\documentclass[11pt]{article}                     % onecolumn (standard format)

\usepackage{graphicx}
\usepackage{epsfig}
\usepackage{amsmath}
\usepackage{mathrsfs}
\usepackage{amssymb}
\usepackage[authoryear]{natbib}
\usepackage[amsmath,thmmarks]{ntheorem}
\usepackage{algorithm,algorithmic}
\usepackage{url}
\usepackage[margin=1.2in,paperwidth=8.5in,paperheight=11in]{geometry}
\usepackage{rotating}
\usepackage{setspace} 
\usepackage{stmaryrd}
\renewcommand{\P}{\mathbb{P}}
\newcommand{\DFT}{\mathrm{DFT}}
\newcommand{\IDFT}{\mathrm{IDFT}}
\newcommand{\MVN}{\mathrm{MVN}}
\newcommand{\ci}{\perp\!\!\!\perp}
\newcommand{\Sigmaext}{\Sigma_{\text{ext}}}
\newcommand{\Qext}{Q_{\text{ext}}}
\newcommand{\tildeSigmaext}{\tilde{\Sigma}_{\text{ext}}}
\newcommand{\tildeQext}{\tilde{Q}_{\text{ext}}}
\newcommand{\thetaopt}{\theta_{\text{opt}}}

\newcommand{\Yext}{Y^{\text{ext}}}
\newcommand{\Gext}{G^{\text{ext}}}

\newcommand{\inv}{^{-1}}
\newcommand{\N}{\mathrm{N}}

\newcommand{\rmd}{\mathrm{d}}

\newcommand{\E}{\mathbb{E}}

\newcommand{\real}{\mathbb{R}}
\newcommand{\cov}{\mathrm{cov}}

\newcommand{\indic}{\mathbb{I}}
\newcommand{\I}{\mathrm{i}}

\newlength{\descwid}
\begin{document}

\title{INLA or MCMC? A Tutorial and Comparative Evaluation for Spatial Prediction in log-Gaussian Cox Processes}

\author{Benjamin M Taylor and Peter J Diggle}

\maketitle

\begin{abstract}
We investigate two options for performing Bayesian inference on spatial log-Gaussian Cox processes assuming a spatially continuous latent field: Markov chain Monte Carlo (MCMC) and the integrated nested Laplace approximation (INLA). We first describe the device of approximating a spatially continuous Gaussian field by a Gaussian Markov random field on a discrete lattice, and present a simulation study showing that, with careful choice of parameter values, small neighbourhood sizes can give excellent approximations. We then introduce the spatial log-Gaussian Cox process and describe MCMC and INLA methods for spatial prediction within this model class. We report the results of a simulation study in which we compare MALA and the technique of approximating the continuous latent field by a discrete one, followed by approximate Bayesian inference via INLA over a selection of 18 simulated scenarios. The results question the notion that the latter technique is both significantly faster and more  robust than MCMC in this setting; 100,000 iterations of the MALA algorithm running in 20 minutes on a desktop PC delivered greater predictive accuracy than the default \verb=INLA= strategy, which ran in 4 minutes and gave comparative performance to the full Laplace approximation which ran in 39 minutes.
\end{abstract}

\section{Introduction}

The primary aim of this article is to provide an objective comparison of two methods for Bayesian Inference in
the spatial log-Gaussian Cox process: a relatively slow but 
asymptotically exact method, Markov chain Monte Carlo (MCMC); and a faster but approximate method, the integrated nested Laplace
approximation (INLA). The secondary aim is to provide a tutorial in some of the technical aspects involved with computation and
inference for this class of models.

The log-Gaussian Cox process is but one of a number of possible 
model classes
that we could have used as the basis for a comparative 
evaluation of MCMC
and INLA methods. 
Our specific
motivation for focusing on this model is its use
in spatial epidemiology, and specifically in health
surveillance applications, where interest is in the predictive probability that the relative risk of disease at a certain spatial
location exceeds a threshold set by public health experts. For an example in the spatio-temporal setting see \cite{diggle2005}.
There, the data consisted 
of locations of incident cases each day, i.e.\ a spatio-temporal
point process, and the Cox process was used to represent
spatio-temporal variation in risk as a product of 
deterministic and stochastic terms representing, respectively, 
known risk-factors and unexplained variation, prediction of which  
was the primary goal. A high predictive probability that risk 
in a particular area
exceeds the pre-declared threshold
may activate a
costly public health intervention, hence it is important
that such
predictive probability statements are as accurate as possible.
For this reason we will compare MCMC and INLA using a metric that directly measures their ability to make 
accurate predictive
probability statements. 

MCMC methods have made an enormous impact on statistical practice
by making Bayesian inference tractable for complex statistical
models, including models
whose specification includes a latent Gaussian process.
However, they can be computationally
burdensome and, more importantly, their inferential validity rests on the
convergence of a Markov chain to its equilibrium distribution,
which can be difficult to verify empirically. 
INLA \citep{rue2009} is a recently
developed competitor to MCMC methods. By using a combination of analytical
approximation and numerical integration rather than Monte Carlo simulation, INLA circumvents the
convergence issues that arise with MCMC methods, 
and typically leads to quicker computation. However,
the price paid is that the analytic approximations
potentially introduce errors
in the calculation of  posterior probabilities. 
The goal of the simulation study in this paper is to assess the 
trade-off between faster computation and errors of approximation.

Our focus is on predicting (functions of) a latent spatially
continuous Gaussian process ${\cal Y}$,
which we approximate by a Gaussian field, $Y$, on a 
finely spaced, regular square grid of points on the plane. 
This is in contrast to the methods discussed in 
\cite{lindgren2011}, in which a representation is constructed on a triangulation of a set of irregularly spaced points. 
\cite{lindgren2011} assume that ${\cal Y}$
has Mat\'ern second order structure, that is for $u,v\in\real^2$,
\begin{equation*}
   \cov(u,v) = \frac{\sigma^2}{\Gamma(\nu)2^{\nu-1}}(\kappa||v-u||)^\nu K_\nu(\kappa||v-u||),
\end{equation*}
where $\nu,\kappa>0$ and $K_\nu$ is a modified Bessel function of the second kind.
The major advantage of their approach is its low
computational cost: for any choice of the admissible 
parameters of the covariance function, they are able 
to compute the precision matrix of the GMRF approximation in $O(n)$ time, where $n$ is the number of triangulation points. 
We prefer to retain greater flexibility in choosing an appropriate model for the covariance, allowing data and scientific
knowledge to inform this choice. For this reason, we focus instead on the method described in Chapter 5 of \cite{rueheld2005},
which allows effectively any covariance model to be fitted to the data.

In Section \ref{sect:gmrfapprox}, 
we discuss the approximation of a spatially
continuous Gaussian process
by a Gaussian Markov random field,
 and give the results
 of a simulation study detailing the effectiveness of this procedure. 
 In Section \ref{sect:lgcpmodel}, 
 we describe
 the spatial log-Gaussian Cox process. Sections \ref{sect:MCMC} and \ref{sect:inlaapprox} give details of the MCMC and INLA
 methods, respectively.
 Section \ref{sect:simulationresults} summarises the findings. 
 Section \ref{sect:discussion} is a concluding discussion. 
 Throughout the article,
 we use  $\pi$ to denote a generic probability density function.

All of the methods discussed in the article are
implemented in the  R package \verb=lgcp=; see \cite{taylor2011a}.

\section{Spatially Continuous Gaussian processes and their Approximation by Gaussian Markov Random Fields}
\label{sect:gmrfapprox}

In this section, we consider how to approximate a spatially continuous Gaussian process by a GMRF and, via simulation, how good
such an approximation is. Note that a discussion of this topic is given in Chapter 5 of \cite{rueheld2005}. To begin, we introduce
some more general concepts.

\subsection{Theory}

A spatially continuous Gaussian process, $\mathcal Y$, is a real-valued continuous Gaussian process on the plane, $\real^2$. This
means that $\mathcal Y$ is a continuous function from $\real^2\rightarrow\real$ with the property that for any finite collection
of locations, $\{s_i\}_{i=1}^n$, the joint distribution of the random variables representing the value of the process at each of
the locations, $\{\mathcal{Y}(s_i)\}$, is multivariate Gaussian. $\mathcal{Y}$ is called {\it strictly stationary} if
$\E(\mathcal{Y}(s_i))=\alpha$ for some $\alpha\in\real$ and any spatial location $s_i$ and {\it strictly second-order stationary and isotropic}, if the covariance between $\mathcal{Y}(s_i)$ and $\mathcal{Y}(s_j)$ only depends on the Euclidean distance between $s_i$
and $s_j$, denoted by $||s_i-s_j||$ \citep{hoss2010}. The covariance between
$u,v\in\real^2$ will be assumed to have the form,
\begin{equation*}
   \cov(u,v) = \sigma^2r(||v-u||/\phi),
\end{equation*}
where $r$ is a standard isotropic correlation function: for example a Mat\'ern function. The parameter
$\sigma$ dictates the point-wise variability of the field, whilst the scale parameter $\phi$ governs the rate of decay of the correlation in
space. In what follows, an italic Roman, $Y$, will be used to denote the values of $\mathcal{Y}$ at a finite set of locations
in space. We say that $Y$ {\it represents} the process $\mathcal{Y}$. 

A Gaussian Markov random field is a collection of random variables, $\tilde Y=\{\tilde Y_1,\ldots,\tilde Y_k\}$, that have a
multivariate Gaussian distribution, $\tilde Y\sim\MVN(\tilde\mu,\tilde Q\inv)$, where for any $j$,
\begin{equation*}
   [\tilde Y_j|\tilde Y_{-j}] = [\tilde Y_j|\text{the neighbours of $\tilde Y_j$}],
\end{equation*}
where $\tilde Y_{-j}$ denotes $\{\tilde Y_1,\ldots,\tilde Y_{j-1},\tilde Y_{j+1},\ldots,\tilde Y_{k}\}$ and $[\,\cdot\,]$ means `the distribution of'. We use $Y$ $\tilde Y$ to distinguish between respectively the Gaussian field and the Gaussian Markov random field representation of a process $\mathcal{Y}$ at the same finite set of locations. The {\it neighbours}
of $j$, ${\cal N}_j$, are usually a much smaller subset of $\tilde Y$;
all other elements of $\tilde Y$ are conditionally independent of $\tilde Y_j$, given ${\cal N}_j$. The pattern of conditional
independence is evident in the precision matrix, $\tilde Q$: Theorem 2.2 in \cite{rueheld2005} states that,
\begin{equation*}
   \tilde Y_i\ci \tilde Y_j|{\cal N}_i \cup {\cal N}_j \iff \tilde Q_{ij}=0.
\end{equation*}
In the case that the neighbourhoods of each element are very small subsets of $\tilde Y$, the matrix $\tilde Q$ is sparse. This
allows otherwise computationally prohibitive operations, such as matrix inversion, to be implemented with fast algorithms.

To simplify matters, consider a square observation window, $W$, on which a spatially continuous Gaussian process is represented by
a finite collection of random variables $Y=\{Y_{ij}\}_{i,j=1}^M$ spaced on a regular square grid , $G=\{G_{ij}\}_{i,j=1}^M$, where the $G_{ij}$ are the
centroids of grid cells, that cover $W$. For computational reasons to be explained, we assume that $M=2^m$ for some positive
integer $m$. In order to obtain a representation of the {\it stationary} second order structure of the process, it is necessary
to extend this grid, typically to a grid of size $2M\times2M$, which is wrapped on a torus. This action gives rise to the notion of a
toroidal distance metric, by which is meant the minimum distance between two points, travelling either directly (e.g.\ if the points
were very close together on the torus) or around the minor and/or major radii. A precise definition is given in
\cite{moller1998}.

Let $\Yext=\{\Yext_{ij}\}_{i,j=1}^{2M}$ be random variables at grid locations $\Gext=\{\Gext_{ij}\}_{i,j=1}^{2M}$ on
the extended space. Note that in cases where the spatial decay parameter, $\phi$, is quite quite large compared with the size of
$W$, then in order to obtain a valid covariance structure, the grid may have to be extended further, e.g.\ onto a $4M\times4M$ toroidal
grid, see for example \cite{moller1998}. The covariance matrix, $\Sigmaext$, of the discrete field $\Yext$ on the extended grid is
typically massive, dense and with a dense inverse, $\Qext$. As an example, for a $128\times128$ grid in the extended space, the
covariance matrix has dimension $16384\times16384$: the storage and manipulation of such matrices under ordinary circumstances is
not computationally feasible on a desktop PC. However, in the extended space, $\Sigmaext$ is
block-circulant (see below) and symmetric positive definite (SPD) with a block circulant SPD inverse, $\Qext$. The symmetry
induced in the covariance matrix by extending the grid and wrapping it on a torus means that each entry of $\Sigmaext$ is one of
exactly $(2M)^2=128^2=16,834$ elements, instead of a possible $16,384^2=268,435,456$ different elements. Furthermore, matrix computation in the extended space is greatly aided by using the discrete Fourier transform (DFT), which is why the focus of this article is on grids of dimension $2^m\times2^m$. In fact, it is possible to drop this assumption, at the price
of reduced speed in the $\DFT$. Algorithms are available to construct optimised computational plans for implementing the $\DFT$ on
other grid sizes, for example the FFTW library \citep{frigo2011} and an R wrapper library \citep{krey2011}. In what follows, $\DFT$
and $\IDFT$ will denote, respectively, the discrete Fourier and discrete inverse-Fourier Transforms; as an abuse of notation, the
same abbreviations will be used for the 1- and 2-dimensional transforms, with the choice being context dependent \citep{wood1994}.

A full discussion of why and how the DFT is used in matrix computations on block circulant matrices is given in Chapter 2 of
\cite{rueheld2005}, but for completeness a very brief summary follows. An $n\times n$ matrix $A$ is said to be {\it circulant}
if it has the following structure:
\begin{equation*}
   A=\left[\begin{array}{cccc}
      a_0 & a_1 & \cdots & a_{n-1}\\
      a_{n-1} & a_0 & \cdots & a_{n-2}\\
      \vdots & \vdots & & \vdots\\
      a_1 & a_2 & \cdots & a_0
   \end{array}\right],
\end{equation*}
where $a_i$ belong to a field, for example the real numbers. The ordered set of elements, $\tilde a=\{a_j\}_{j=1}^n$, is called
the base of $A$. A real $nm\times nm$ block circulant matrix $C$ is one with the following structure:
\begin{equation*}
   C=\left[\begin{array}{cccc}
      A_0 & A_1 & \cdots & A_{m-1}\\
      A_{m-1} & A_0 & \cdots & A_{m-2}\\
      \vdots & \vdots & & \vdots\\
      A_1 & A_2 & \cdots & A_0
   \end{array}\right],
\end{equation*}
where each $A_i$ is a circulant matrix with base $\tilde a_i=\{a_{ij}\}_{j=0}^{n-1}$. The matrix
\begin{equation*}
   \tilde c=\left[\begin{array}{cccc}
      a_{00} & a_{10} & \cdots & a_{(m-1)0}\\
      a_{01} & a_{11} & \cdots & a_{(m-1)1}\\
      \vdots & \vdots & & \vdots\\
      a_{0(n-1)} & a_{1(n-1)} & \cdots & a_{(m-1)(n-1)}
   \end{array}\right]
\end{equation*}
is known as the base matrix of $C$.

The eigenvectors of a circulant matrix $A$ (as defined above) can be written as
\begin{equation*}
   E_j = \sum_{i=0}^{n-1}a_i\exp\{-2\pi\I ij/n\}.
\end{equation*}
The complete set of eigenvectors are stored in columns of the (unitary) discrete Fourier transform matrix,
\begin{equation*}
   F = \left[\begin{array}{ccccc}1&1&1&\cdots&1\\ 1&\omega&\omega^{2}&\cdots&\omega^{n-1} \\
1&\omega^{2}&\omega^{4}&\cdots&\omega^{2(n-1)} \\ \vdots&\vdots&\vdots&\ddots&\vdots\\
1&\omega^{n-1}&\omega^{2(n-1)}&\cdots&\omega^{(n-1)(n-1)}\end{array}\right],
\end{equation*}
where $\omega=\exp\{2\pi\I/n\}$. Now, let $E$ be a diagonal matrix with the eigenvalues $\{E_j\}_{j=0}^{n-1}$ on the leading
diagonal. By expanding the matrix product analytically it is straightforward to verify that the matrix $A$ has spectral
decomposition, $A = FEF^H$, where the superscript $H$ denotes the conjugate transpose. The most useful aspect of the matrix $F$ is that matrix-vector products $Fv$ and
$F^Hv$ are available directly as the $\DFT$ and $\IDFT$, respectively, of the vector $v$. The vector $E$ is available as $E = \sqrt{n}F\tilde a$ and matrix square roots are computed from expressions such as $A^{1/2} = FE^{1/2}F^H$. These results massively simplify computation with ordinary circulant matrices; furthermore, the theory extends to block circulant
matrices, where the 2-dimensional $\DFT$ and $\IDFT$ are used.

In what is to follow the `full covariance matrix' will mean the covariance matrix of $\Yext$, i.e.\ $\Sigmaext$. For the application
in mind, namely the spatial log-Gaussian Cox process, it has been argued that Markov chain Monte Carlo using the full covariance
matrix is inefficient \citep{rue2009,simpson2011}. The suggested alternative is to use the integrated nested Laplace
approximation to perform approximate Bayesian inference with a sparse GMRF approximation to the full covariance matrix so as to reduce
computational cost. We now discuss the construction of such a GMRF approximation.

For the full covariance matrix defined by an arbitrary choice of correlation function, the dependence structure is `dense': there is no sparse conditional independence structure -- the
precision matrix $\Qext$ is a dense matrix. A GMRF approximation to $\Sigmaext$ is constructed by parametrising the precision
matrix, $\tildeQext\equiv\tildeQext(\theta)$, and choosing $\thetaopt$ to be such that
$\tildeQext(\thetaopt)\inv=\tildeSigmaext(\thetaopt)$ is `as close as possible' to $\Sigmaext$.
\begin{figure}[htbp]
   \centering
   \includegraphics[width=0.6\textwidth,height=0.3\textwidth]{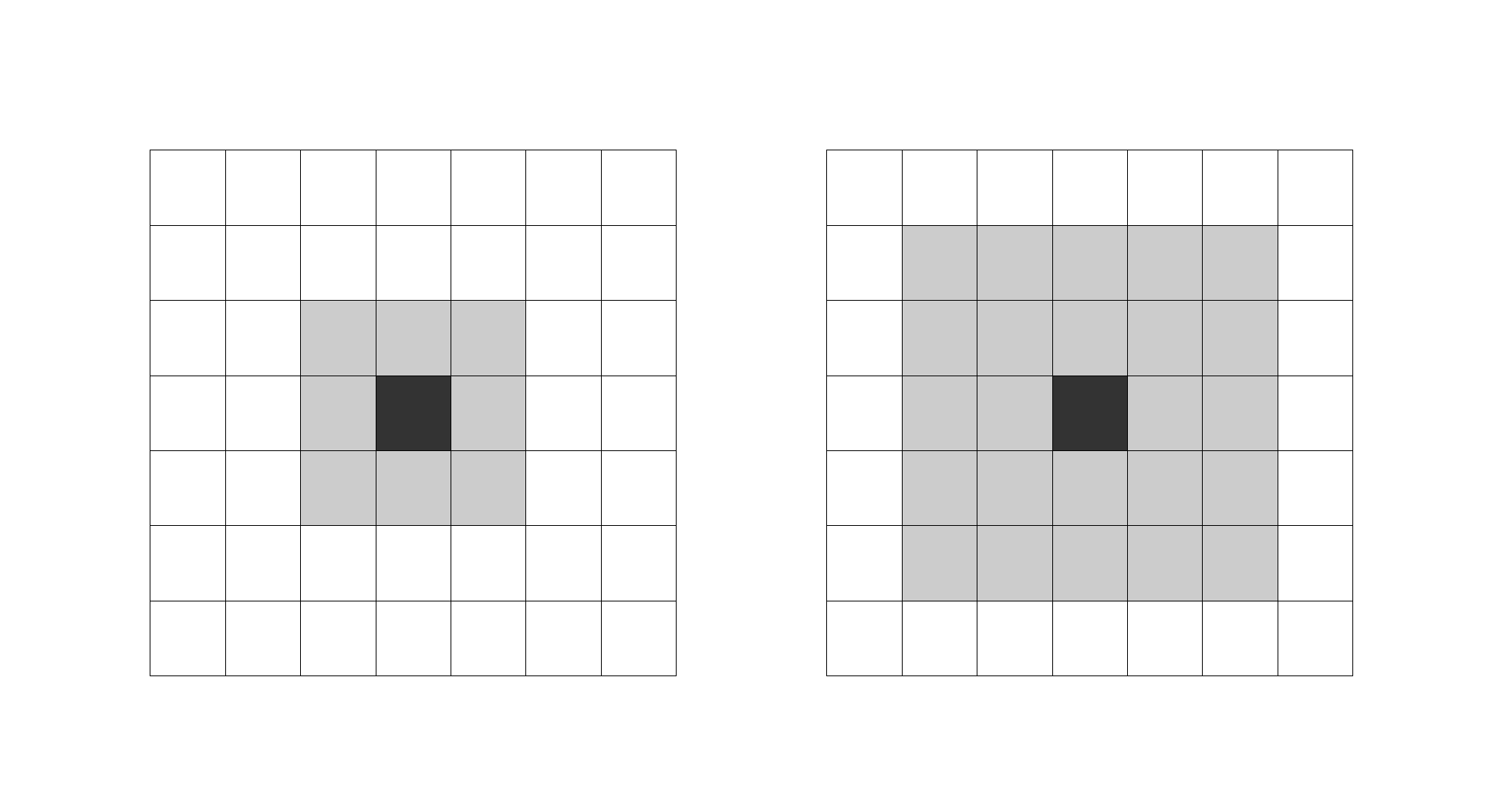}
   \caption{\label{fig:neigh} Illustrating the choice of 1- and 2-neighbourhoods, respectively the left and right hand plots. The
black square is the grid cell of interest and the grey cells are those specified to be the neighbours in the chosen level of
dependence.}
\end{figure}
The parametrisation considered here is similar to that presented in Chapter 5 of \cite{rueheld2005}, in which the neighbourhood of
a grid cell consists of all the cells in the box surrounding the cell of interest, up to a certain distance away. Figure \ref{fig:neigh}
illustrates what will be referred to in this article as 1- and 2-neighbourhoods: the box obtained by specifying respectively dependence on all
cells in the box a distance of up to 1 and up to 2 cells both vertically and horizontally around the cell of interest. For the
1-neighbourhoods, $\theta$ has 3 elements, corresponding to the dependences between cells a distance 0 apart, between directly adjacent
cells and between diagonally adjacent cells. In a similar way, for the 2-neighbourhood dependence structure, $\theta$ has six elements.

Let $\varsigma$ be the base matrix of $\Sigmaext$ and let $\hat\varsigma(\theta)$ be the base of the inverse of $\tildeQext$, with
base matrix $\tilde\psi(\theta)$. Note that, $\hat\varsigma(\theta)$ can be computed from $\tilde\psi(\theta)$ using the 2-dimensional
discrete Fourier transform,
\begin{equation*}
   \hat\varsigma(\theta)=\frac{1}{\text{length}\{\text{vec}[\tilde\psi(\theta)]\}}\IDFT(\DFT(\tilde\psi(\theta))\varowedge(-1)),
\end{equation*}
where $X\varowedge(-1)$ denotes the raising of each element of matrix $X$ to the power $-1$ and $\text{vec}(X)$ is the vector
obtained by stacking the columns of $X$ on top of each other. The optimal $\theta$ is found as
$\min_{\theta}\left\{U(\theta)\right\}$ where,
\begin{equation*}
   U(\theta) = \sum_{ij}w_{ij}(\varsigma_{ij}-\hat\varsigma(\theta)_{ij})^2,
\end{equation*}
where a subscript $ij$ denotes the $(i,j)$-element of the matrix,
\begin{equation*}
   w_{ij} \propto \left\{\begin{array}{ll} 1 & \text{if $ij=(0,0)$} \\ \frac{1+a/d(i,j)}{d(i,j)} & \text{otherwise}\end{array}\right.
\end{equation*}
Here, $d(i,j)$ is the distance from cell $(i,j)$ to
the reference origin $(0,0)$ and $a$ is a constant, set equal to 1 in the experiments below, see \cite{rueheld2005} for a justification of this choice.

Differentiating $U$ with respect to the $k$th component of the parameter, $\theta_k$, gives
\begin{equation*}
   \frac{\partial U(\theta)}{\partial\theta_k} =
-2\sum_{ij}w_{ij}(\varsigma_{ij}-\hat\varsigma(\theta)_{ij})\frac{\partial\hat\varsigma(\theta)_{ij}}{\partial\theta_k},
\end{equation*}
which can also be computed using the DFT, since
\begin{equation*}
   \frac{\partial \varsigma(\theta)}{\partial\theta_k} =
\frac{1}{\text{length}\{\text{vec}[\tilde\psi(\theta)]\}}\IDFT\left\{-[\DFT(\tilde\psi(\theta))\varowedge(-2)]\odot\DFT\left[\frac{\partial}{
\partial \theta_k}\tilde\psi(\theta)\right]\right\}.
\end{equation*}
where $\odot$ denotes element-wise multiplication, and the matrix $\frac{\partial}{\partial \theta_k}\tilde\psi(\theta)$ is a matrix
of 1's where $\theta_k$ appears and 0's otherwise.

With the above ingredients, standard software such as the \verb=optim= function in R can be used to compute optimal parameters. In
particular, the availability of the gradient function enables the user to take advantage of gradient-based optimisation methods
such as the \verb=BFGS= method implemented in \verb=optim=. A sensible starting point for the optimiser is given by the base
matrix of the diagonal precision matrix, $\text{diag}(1/\sigma^2)$.

The reader who is daunted by the prospect of implementing the above functions should note that this has already been done in the
\verb=lgcp= R package \citep{taylor2011a}.

\subsection{Simulation Study}

We performed a simulation study to investigate the ability of the algorithm detailed above to approximate a spatially continuous
Gaussian process, and hence choose an appropriate neighbourhood size for use later in the article (see Section
\ref{sect:simulationresults}).

\begin{figure}[htbp]
   \centering
   \includegraphics[width=0.3\textwidth,height=0.3\textwidth]{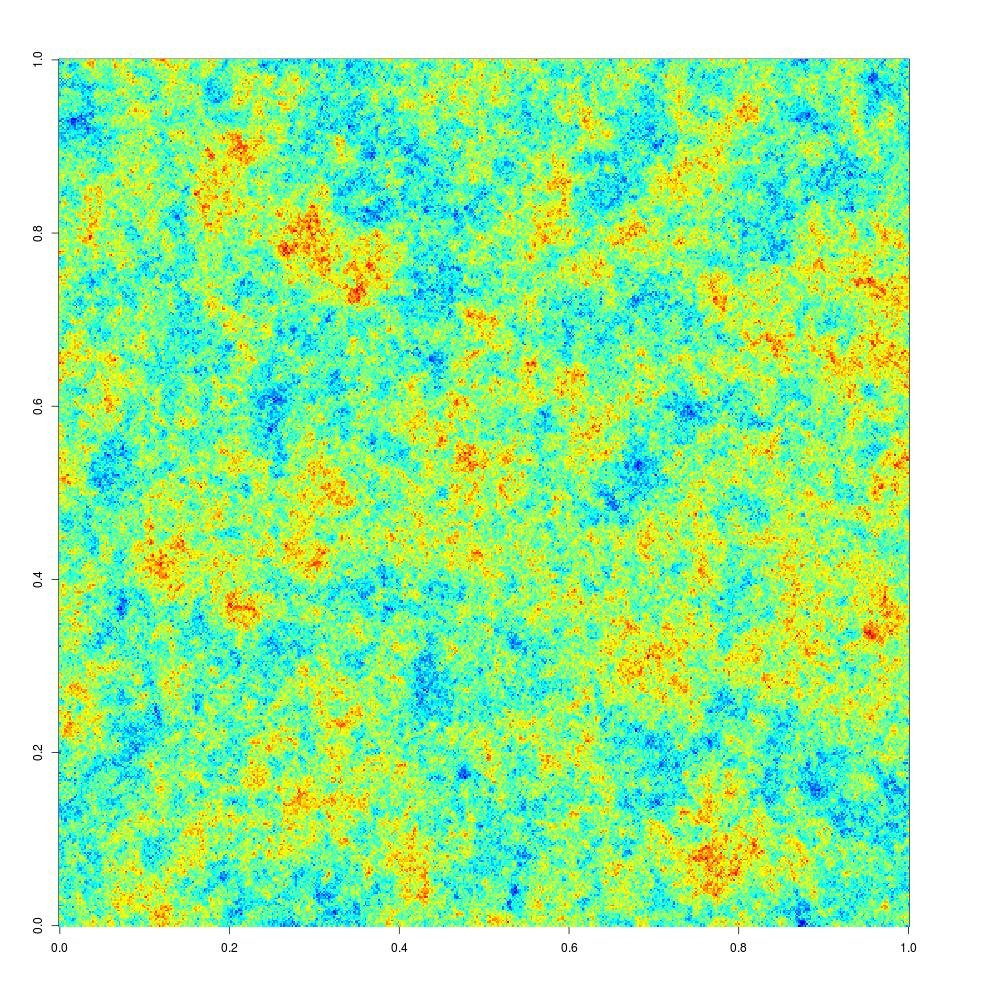}
   \includegraphics[width=0.3\textwidth,height=0.3\textwidth]{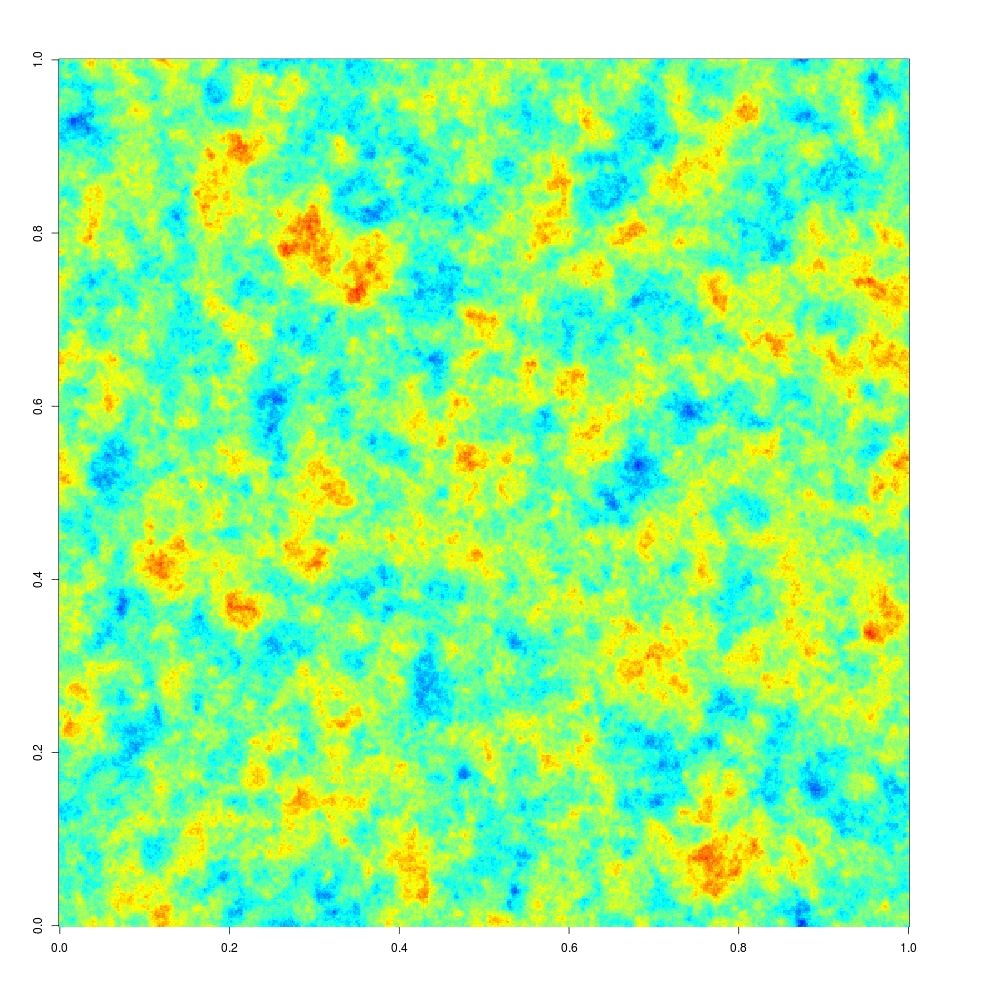}
   \includegraphics[width=0.3\textwidth,height=0.3\textwidth]{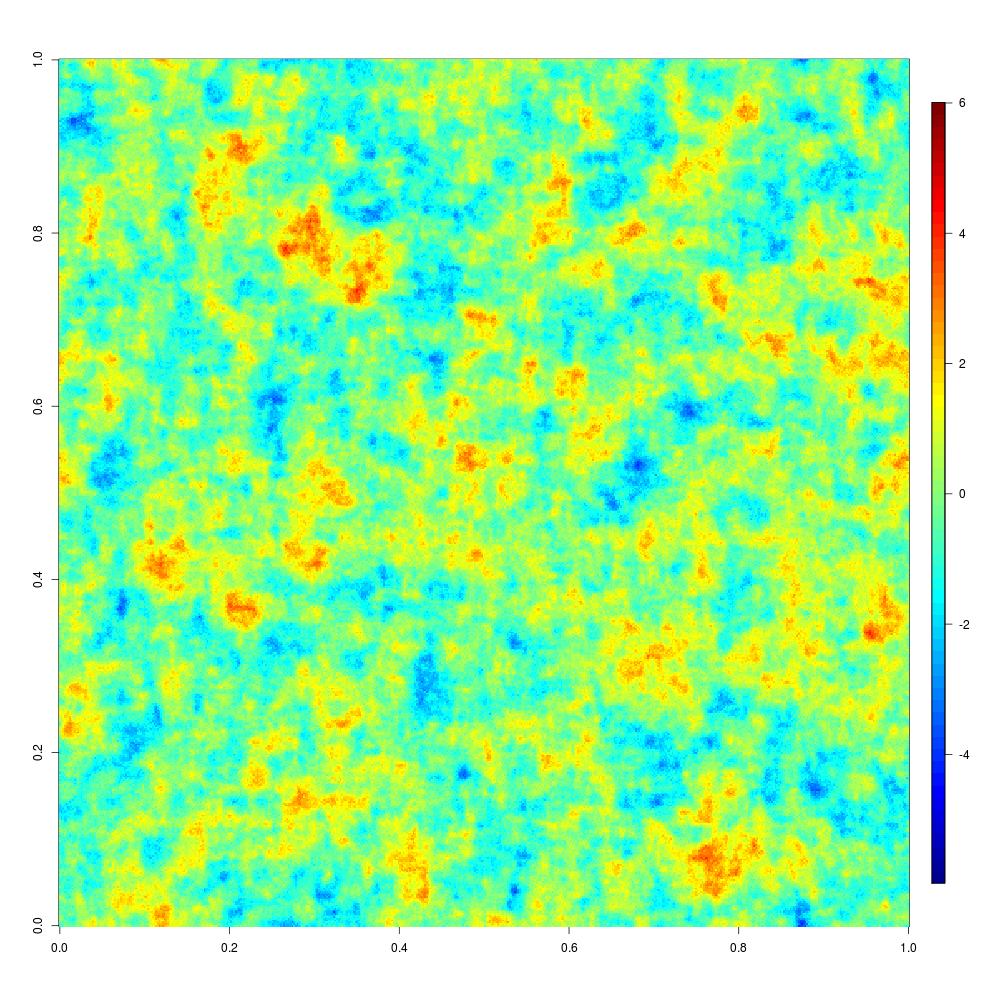}
   \caption{\label{fig:approxfieldsim} Simulated Gaussian fields. Middle plot: the true field with full covariance structure;
left-hand plot: approximate field with a 1-neighbourhood; right-hand plot: approximate field with a 2-neighbourhood. The
simulation took place on a $512\times512$ grid, the optimisation took respectively 64 and 355 seconds to compute the parameters of
the 1- and 2- neighbourhoods. The result from the 2-neighbourhood is virtually indistinguishable to the eye from the true field,
whereas the 1-neighbourhood has a grainy appearance. The respective MSE's were 0.38 and 0.007.}
\end{figure}

Given a set of parameters of the latent field, an observation window and a grid size, it is possible to compute the base matrix of
$\Sigmaext$. With the same inputs and an optimisation step, it is also possible to compute the base matrix of the sparse
representation, $\tildeSigmaext(\thetaopt)$. A measure of the performance of the approximation is given by the mean square error in
simulating Gaussian random variables. Given a vector of standard Gaussian variates, $\Gamma$, a realisation of a Gaussian field
with mean $\mu$ and the correct second order properties is given by $y = \mu + \Sigmaext^{1/2}\Gamma$, whilst an approximation of the field is $\tilde y = \mu + \tildeSigmaext^{1/2}\Gamma$. To compare how well the field has been approximated, an appropriate measure is the integrated mean square error. We estimate this
from a repeated sequence of $n$ independent
realisations of $\Gamma$ as
\begin{equation}
   \text{MSE} = \frac1{nM^2}\sum_{i=1}^n\sum_{j=1}^M\sum_{k=1}^M (y_{ijk}-\tilde y_{ijk})^2,
\end{equation}
where $y_{ijk}$ is the value of the $(j,k)$ cell of $y$ for the $i$th realisation of $\Gamma$ and $M$ is the number of grid cells
in each direction (here, the grid is assumed to be square). Figure \ref{fig:approxfieldsim} is a visualisation of a true field and
two possible sparse approximations using 1- and 2-neighbourhoods.

% Optimiser took 64.03232 
% Optimiser took 355.2584 
% 
% 256 by 256 ext 512 512
% 
% [1] "MSE neighbourhood size 1 0.380888225568048"
%      paste("MSE neighbourhood size 2",mean((trueY-approxY2)^2))
% [1] "MSE neighbourhood size 2 0.00741925603759391"

For the simulation study, the value of $\sigma$ was fixed at 1 and $\phi$ allowed to vary between 0.025 and 0.2 with the unit
square as the observation window. The parameter $\sigma$ was fixed because the computed mean square errors would simply scale linearly with $\sigma^2$. The range of values of $\phi$ was chosen to include a selection of scenarios that might be encountered in practice; the important factor is the size of $\phi$ with respect to
the observation window and the grid. When $\phi$ is small compared to the size of the grid, the cells become approximately independent. It
should be harder to obtain a good approximation to the latent field for larger values of $\phi$, as in this case spatial
dependence can be significant for cells a moderate distance apart on the grid. To reduce the possibility of results depending on
an artefact of the choice of grid size, two different resolutions were used: $128\times128$ and $256\times256$ in the extended
space. For the comparison of $x$ and $\tilde x$, the parameter $\mu$ was set to zero.

\begin{table}
   \footnotesize
    \centering
    \begin{tabular}{c|c|ccc|ccc}
         Ext. Grid & $\phi$ & MSE 1 & MSE 2 & MSE 3 & Bias 1 & Bias 2 & Bias 3 \\ \hline
        128 & 0.025 & 0.086 (0.6) & 0.014 (6.2) & 0.023 (9.7) & $1.179\times10^{-4}$ & $3.86\times10^{-5}$ & $1.424\times10^{-5}$
\\
        128 & 0.05 & 0.174 (1.1) & 0.025 (6.4) & 0.062 (9.8) & $5\times10^{-4}$ & $1.333\times10^{-4}$ & $0.618\times10^{-4}$ \\
        128 & 0.1 & 0.283 (1.1) & 0.016 (6.5) & 0.042 (9.8) & 0.002 & $4.332\times10^{-4}$ & $1.891\times10^{-4}$ \\
        128 & 0.15 & 0.358 (1) & 0.018 (6.7) & 0.03 (9.8) & 0.003 & 0.001 & $4.338\times10^{-4}$ \\
        128 & 0.2 & 0.419 (1.5) & 0.031 (6.8) & 0.064 (9.9) & 0.004 & 0.001 & 0.001  \\ \hline
        256 & 0.025 & 0.173 (5.2) & 0.024 (33.6) & 0.005 (58.5) & $-1.92\times10^{-4}$ & $-0.535\times10^{-4}$ &
$-1.005\times10^{-5}$ \\
        256 & 0.05 & 0.278 (5.6) & 0.024 (35.4) & 0.034 (59.7) & -0.001 & $-1.848\times10^{-4}$ & $-0.703\times10^{-4}$ \\
        256 & 0.1 & 0.389 (7.7) & 0.041 (35.4) & 0.069 (61.5) & -0.002 & -0.001 & $-3.078\times10^{-4}$ \\
        256 & 0.15 & 0.462 (6.6) & 1.677$^\star$ (32.7) & 0.051 (61) & -0.004 & -0.003 & -0.001  \\
        256 & 0.2 & 0.515 (8.5) & 1.826$^{\star\star}$ (27.1) & 0.051 (58.6) & -0.004 & -0.003 & -0.001  \\
    \end{tabular}
    \caption{\label{tab:GMRFapproxtest} Table illustrating the ability of a GMRF to approximate a GF on a unit square observation
window. `Ext. Grid' is the size of the DFT grid used, giving respective output grid sizes of $64\times64$ and $128\times128$; `MSE
1--3' denotes the mean square error for respective neighbourhood sizes 1--3, with computation time in seconds in parenthesis; and
`Bias 1--3' is the mean bias. Optimisation was performed using the `BFGS' method in R's optim command. Using the Nelder-Mead
simplex method, the two exceptional values, marked $^\star$ and $^{\star\star}$ had improved MSE's of 0.284 and 0.378. On these
occasions, it appeared that the BFGS optimiser had converged to a sub-optimal set of parameters.}
\end{table}

Table \ref{tab:GMRFapproxtest} presents the results of the simulation study. There was a significant
improvement in the approximation using a neighbourhood size of 2 compared with a neighbourhood size of 1. The 3-neighbourhood
approximations did not appear to improve the quality of the estimated latent field, but were more robust: for the 2-neighbourhood
results, the \verb=BFGS= option within \verb=optim= appeared to converge to a local sub-optimal value on two of the occasions, but improved results were
obtained by the \verb=Nelder-Mead= simplex option in these cases.

The main conclusions from this study are as follows: (1) it is possible to compute a very good sparse approximation to a given
Gaussian field for sensible values of $\phi$ compared with the size of the observation window; (2) the optimisation time is quick:
around a minute for $256\times256$ grids; (3) the optimisation step is not always robust, so in practice, building in a simulation
step such as described above, i.e.\ computing $x$ and $\tilde x$ for a range of $\Gamma$, is worthwhile in order to evaluate the
approximation; finally (4) for the purposes of the simulation study in Section \ref{sect:simulationresults}, the 2-neighbourhoods
should be sufficiently accurate, because the values of $\phi$ considered in Section \ref{sect:simulationresults} are sufficiently small
compared with the observation window.

\section{The Spatial Log-Gaussian Cox Process}
\label{sect:lgcpmodel}

Let $W\subset\real^2$ be an observation window in space. Events occur at spatial positions $s\in W$ according to an inhomogeneous
spatial Cox process with intensity function $R(s)$. Conditional on $R$, the
number of events, $X_{S}$, occurring in any $S \subseteq W$ is Poisson distributed \citep{moller1998},
\begin{equation}\label{eqn:themodel}
   X_{S} \sim \text{Poisson}\left\{\int_S R(s)\rmd s\right\}.
\end{equation}
Following \cite{diggle2005}, we decompose the intensity multiplicatively as,
\begin{equation}\label{eq:multiplicative}
   R(s) = \mu\lambda(s)\exp\{\mathcal Y(s)\}.
\end{equation}
In Equation \ref{eq:multiplicative}, the
{\it fixed spatial component}, $\lambda:\real^2\mapsto\real_{\geq 0}$, is a
known function, proportional to the average intensity of the process at each point in space
and scaled so that
\begin{equation}\label{eqn:intlambdas}
   \int_W\lambda(s)\rmd s=1,
\end{equation}
whilst $\mu$ is the expected number of events in $W$. The function $\mathcal Y$ captures the residual variation and is a
continuous spatial Gaussian process. The components $\lambda$ and $\mu$ define the expected behaviour of the point process, whereas $\mathcal Y$ determines the
residual variation. When $\lambda$ and $\mu$ are known, it is this residual variation that is of statistical interest: inference about
$\mathcal Y$ gives information on extraneous spatial clustering of events, with many applications in ecology and epidemiology
\citep{moller1998,diggle2005,rue2009,simpson2011}.

In this article, $\mathcal Y$, is second order stationary with
minimally-parametrised covariance function,
\begin{equation}\label{eq:cov}
\cov[\mathcal Y(s_1),\mathcal Y(s_2)] =
\sigma^2r(||s_2-s_1||/\phi)\},
\end{equation}
where $\sigma$ and $\phi$ are known parameters.  The parameter $\sigma$ scales the log-intensity, 
whilst the parameter $\phi$ governs the rate 
at which the correlation function decreases in space. The mean of the process $\mathcal Y$ is set equal to $-\sigma^2/2$
so as to give $\E[\exp\{\mathcal Y\}]=1$.

Following  \cite{moller1998},
\cite{brix2001} and \cite{diggle2005} we will use
a discretised version, $Y$, of the above model,
 defined on a regular square grid. Strictly,
 observations $X$ arising from the model are 
 then cell counts on the grid, the intensity of the process being treated as constant within each cell. In practice, the aim is to
make the
 lattice spacing sufficiently fine that cell-counts greater than 1 are rare
 and the error introduced by
 piece-wise constant approximation to the intensity is negligible.
 
 Recall that the discretised $Y$ is a finite collection of random variables, so the properties of $\mathcal Y$ imply that $Y$ has
a multivariate Gaussian density with approximate covariance matrix $\Sigma$. The elements of $\Sigma$ are calculated by evaluating
Equation \ref{eq:cov} at the centroids of the spatial grid cells. 

The next two sections present alternative methods for predictive inference of $Y$ conditional on the observed data $X$.

\section{Markov Chain Monte Carlo}
\label{sect:MCMC}

The main advantage of the MCMC approach to inference for this model is that the `full
covariance matrix' of $Y$ is used (compare with Section \ref{sect:inlaapprox}). Predictive inference about $Y$ requires samples
from the conditional distribution of the latent field, $Y$, given the observations, $X$, namely
\begin{equation}\label{eqn:jointdens}
   \pi(Y|X) \propto \pi(X|Y)\pi(Y),
\end{equation}
In order to evaluate $\pi(Y)$ in Equation \ref{eqn:jointdens}, the parameters of the process $Y$ must either be known or estimated
from the data. Estimation of $\sigma$ and $\phi$ can be achieved either in a Bayesian or likelihood-based framework, or by one of a number of more {\it
ad hoc} methods e.g.\ \cite{brix2001} and \cite{diggle2005}. These ad-hoc methods are described and implemented in the R package
\verb=lgcp=, see \cite{taylor2011a}.

Markov chain Monte Carlo methods work by drawing a sequence of dependent samples from a target density of interest in situations
where it is not possible to draw independent samples directly (e.g.\ via inversion of the cumulative distribution function). In the limit as the
number of draws tends to infinity, the resulting output behaves as if it were a sample 
from the density of interest. In practice, if the chain appears to have
converged to its stationary distribution
 and is mixing well, that is to say that the autocorrelation of the generated sample is low, then reliable inference can be
can be made with relatively short runs. However, for non-trivial applications, the chain often mixes poorly, and it is difficult to
tell whether stationarity has been achieved without prohibitively long runs to double-check that this is the case. Reviews of MCMC
methodology include \cite{MCMCiP} and \cite{gamermanlopes}.

The most well known MCMC method, the Metropolis Hastings (MH) algorithm \citep{metropolis1953,hastings1970} is ``arguably the most
successful and influential Monte Carlo [technique]'' \citep{girolami2011}; along with Gibbs sampling, it is probably the most
frequently employed method in the literature. Let $\pi$ be the target density. Given a current state in the chain, $Y$, a new
state, $Y'$, is proposed from any probability density, $q(Y,Y')= \P(Y'|Y)$, and accepted with probability,
\begin{equation}\label{eqn:mhacc}
\alpha(Y,Y') =
\min\left\{1,\frac{\pi(Y'|X)}{\pi(Y|X)}\frac{q(Y',Y)}{q(Y,Y')}\right\}.
\end{equation}
A crucial step in designing an effective sampling regime is the choice of proposal kernel $q$. The word `choice' both describes
the specific probability density function associated with $q$, as well as any tuning parameters, $h$, that are associated with it.
A very simple example is a random-walk kernel, where $q(Y,Y')\sim\N(Y,h^2)$. For inference with the log-Gaussian Cox process, this article will focus on a more sophisticated, but well-studied version
of the Metropolis-Hastings algorithm, the Metropolis-Adjusted Langevin algorithm.

Following \cite{moller1998}, Monte Carlo simulation from $\pi(Y|X)$ is made more efficient by working with a linear transformation
of $\Yext$. Writing $\Gamma=\Sigmaext^{-1/2}(\Yext-\sigma^2/2)$, the target of interest is given by,
\begin{equation}\label{eqn:target}
   \pi(\Gamma|X) \propto \pi(X|\Yext)\pi(\Gamma),
\end{equation}
When the gradient of the transition density can also be written down explicitly, a natural and efficient MCMC method for sampling
from the predictive density of interest (Equation \ref{eqn:target}), is a Metropolis-Hastings algorithm with a Langevin-type
proposal \citep{roberts1996,moller1998},
\begin{equation*}
   q(\Gamma,\Gamma') = \N\left[\Gamma';\Gamma + \frac12\nabla\log\{\pi(\Gamma|X)\},h^2\indic\right],
\end{equation*}
where $\N(y;m,v)$ denotes a Gaussian density with mean $m$ and variance $v$ evaluated at $y$, $\indic$ is the identity matrix and
$h>0$ is a scaling parameter \citep{metropolis1953,hastings1970}. 

Various theoretical results exist concerning the optimal acceptance probability of the MALA (Metropolis-Adjusted Langevin
Algorithm); see \cite{roberts1998} and \cite{roberts2001}. In practical applications, the target acceptance probability is often
set to 0.574, which would be approximately optimal for a Gaussian target as the dimension of the problem tends to infinity. An
algorithm for the automatic choice of $h$ so that this acceptance probability is achieved without disturbing the ergodic property
of the chain is detailed in \cite{andrieu2008}, and is implemented in the \verb=lgcp= R package \citep{taylor2011a}.

In our case,
the log of the target density (\ref{eqn:target}) is given by
\begin{equation*}%%\label{eqn:logtarget}
   \log\{\pi(\gamma|X)\} = \text{constant} - \frac12||\gamma||^2 + \sum_{s\in S} \left\{Y(s)X(s) - \mu
C_A\lambda(s)\exp(Y(s))\right\},
\end{equation*}
where $S$ is the set of grid cells in the observation window, $C_A$ is the area of the individual grid cells and $X(s)$ is the
number of events in cell $s$. The gradient can be computed using the $\DFT$,
\begin{equation*}
   \nabla\log\{\pi(\gamma|X)\}(s) = -\gamma(s) + \frac{1}{\text{length}\{\text{vec}[\Yext]\}}\left(\DFT\{E^{1/2}\odot\IDFT[X - \mu
C_A\lambda\exp(\Yext)]\}\right)(s),
\end{equation*}
where in this case, $s$ is a cell in the extended grid.

\section{The Integrated Nested Laplace Approximation}
\label{sect:inlaapprox}

Further details on the material in this section are given in \cite{rue2009}. The Gaussian Markov random field/integrated
nested Laplace approximation (GMRF/INLA) approach to inference for spatial log-Gaussian Cox processes relies on two different
approximations. Firstly, an approximation to the full covariance matrix of $Y$ is obtained, yielding substantial computational
benefits (see Section \ref{sect:gmrfapprox}). Then, this approximate covariance takes the place of the full covariance in the
model, and Bayesian inference is performed using the integrated nested Laplace approximation.

The integrated nested Laplace approximation delivers approximate inference for the posterior marginals of the latent field given the
observed data and also for parameters of the latent field. Let $\vartheta$ be hyperparameters of the latent field $Y$, noting that
these are different to the parameters $(\theta,\phi)$ mentioned above. INLA is based on standard results for marginal and
conditional densities:
\begin{eqnarray*}
   \pi(Y(s)|X) &=& \int\pi(Y(s)|\vartheta,X)\pi(\vartheta|X)\rmd\vartheta,\\
   \pi(\vartheta_k|X) &=& \int\pi(\vartheta|X)\rmd\vartheta_{-k},
\end{eqnarray*}
Where, conditional on the observed data, $\pi(Y(s)|X)$ is the posterior marginal density of the latent field at cell $s$  and
$\pi(\vartheta_k|X)$ is the marginal posterior density of the $k$th component of $\vartheta$. With INLA, these exact relations are
replaced by the approximations,
\begin{eqnarray*}
   \tilde\pi(Y(s)|X) &=& \int\tilde\pi(Y(s)|\vartheta,X)\tilde\pi(\vartheta|X)\rmd\vartheta,\\
   \tilde\pi(\vartheta_k|X) &=& \int\tilde\pi(\vartheta|X)\rmd\vartheta_{-k},
\end{eqnarray*}
where $\tilde\pi$ denotes an approximation to a probability density function (pdf). The approximation of the joint density of the
hyperparameters $\vartheta$ uses
\begin{equation*}
   \tilde\pi(\vartheta|X) \propto \left.\frac{\pi(Y,\vartheta,X)}{\pi_\mathrm{G}(Y|\vartheta,X)}\right|_{Y=Y^\star(\vartheta)},
\end{equation*}
where $\pi_G$ denotes a Gaussian approximation to a density and $Y^\star(\vartheta)$ is the mode of the full conditional for $Y$.
Lastly, the remaining undefined pdf is also approximated by a Gaussian,
\begin{equation*}
   \tilde\pi(Y(s)|\vartheta,X) = \N[Y(s);\mu_s(\vartheta),\sigma_s^2(\vartheta)].
\end{equation*}
To evaluate the marginal posteriors, \cite{rue2009} suggest a clever quadrature scheme over $\vartheta$. Optimised numerical routines for
performing the above computations, and a large suite of other methods not covered in this article, are implemented in the R
\verb=INLA= package available from \url{www.r-inla.org} \citep{rue2009}.

For the simulation study below, Bayesian inference was performed using \verb=INLA='s \verb=generic0= model. The default method for
fitting this model includes a specification for the precision matrix, $\tau\Qext$, where $\tau$ is a hyperparameter with a log-gamma
prior. For the application considered here, the parameter $\tau$ is redundant because, having parametrised and estimated the
precision matrix (as detailed in Section \ref{sect:gmrfapprox}), $\tildeQext$ is a known quantity. The additional
uncertainty induced by the default choice of prior was removed by fixing $\tau=1$. Fixing $\tau$ also considerably speeds up the
time taken to fit the model. Since for the simulations, the mean $\mu$ is known, the model was fitted without an offset (for those familiar with the \verb=INLA= package, \verb=form <- X~ -1 + f(...)=).

% % form <- X~ -1 + f(index,model="generic0", Cmatrix=sbase,hyper=generic0hyper)
% % result <- inla(form,family="poisson",E=E,data=stdata,verbose=inlaverbose,quantiles=q)

\section{Simulation Results}
\label{sect:simulationresults}

\subsection{The Simulated Scenarios}

We simulated a total of 18 scenarios on a unit square observation window. We generated two different fixed spatial components, $\lambda_1(s)$
and $\lambda_2(s)$, by scattering 200 points uniformly on the unit square and fitting a fixed-bandwidth bivariate
Gaussian smoothing kernel to these, with respective standard deviations 0.04 and 0.1.

For each of the fixed spatial components, we used an exponential covariance model with all combinations of $\phi=$0.02, 0.04, 0.6 and $\sigma=$ 0.5, 1, 2. Since the latent field is exponentiated in the model, increasing values of $\sigma$ make predictive inference more challenging for both MALA and INLA. We also hypothesised that larger values of $\phi$ would present a greater challenge for the GMRF approximation step, and hence for the approximate inference step using INLA.

Note that for the full Gaussian field comparison below, we use the word 'INLA' to mean 'a GMRF approximation to the latent field followed by inference via the integrated nested Laplace approximation', this is distinct from the R library \verb=INLA=.

Simulation and prediction took place on an identical grid size of $128\times128$ in the extended space. This enabled a direct
comparison of the true $Y$ used to generate the data, and the predicted $Y|X$ generated from the fitting processes. All computations were carried out on a 3.2GHz Intel(R) Core(TM) i5 desktop PC with 4Gb RAM.

\subsection{Algorithm Parameters}

For inference with MCMC, the sampler was set to run for 100,000 iterations with a 10,000 iteration burn-in and every 90th sample
retained. The choice of 100,000 iterations was primarily for computational reasons: pilot runs had shown that results could be
generated relatively quickly with this choice (in under 20 minutes), whilst the issue of convergence will be discussed in Section
\ref{sect:results}. The chain was initialised with $\Gamma$ set to zero, corresponding to the mean of the field $Y$, i.e.\
$-\sigma^2/2$. For the adaptive MCMC, we used a method introduced by \cite{andrieu2008} that is built into the
\verb=lgcp= package. This uses a Robbins-Munro stochastic approximation to adapt the tuning parameter of the proposal kernel
\citep{robbins1951}. At each iteration of the sampler, the tuning parameter is updated according to the iterative scheme,
\begin{equation*}
   h^{(i+1)} = h^{(i)} + \eta^{(i+1)}(\alpha^{(i)} - \alpha_\text{opt}),
\end{equation*}
where $h{(i)}$ and $\alpha^{(i)}$ are the tuning parameter and acceptance probability at iteration $i$ and $\alpha_\text{opt}$ is
the target acceptance probability. The sequence $\{\eta^{(i)}\}$ is chosen so that $\sum_{i=0}^\infty\eta^{(i)}$ is infinite but
for some $\epsilon>0$, $\sum_{i=0}^\infty\left(\eta^{(i)}\right)^{1+\epsilon}$ is finite. These two conditions ensure that any
value of $h$ can be reached, but in a way that maintains the ergodic behaviour of the chain. The class of sequences used in this
simulation study was $\eta^{(i)} = C/i^\alpha$ where  $\alpha=0.5$ and $C=1$; the scheme was initialised with $h=1$ and set to target $\alpha_\text{opt}=0.574$.

For inference with INLA, we compared a GMRF approximation using neighbourhood sizes of both 1 and 2. The first two methods employ a Gaussian approximation to the target (in \verb=INLA= this corresponds to \verb/strategy="simplified.laplace"/), which will henceforth be referred to as GAIN1 and GAIN2 (`GMRF Approximation followed by INLA'). The third method, GAIN3, employed a neighbourhood size of two and the full Laplace approximation to the target (\verb/strategy="laplace"/). Fitting was achieved with \verb=INLA='s \verb=generic0= model, as discussed in Section \ref{sect:inlaapprox}.

\subsection{Results}
\label{sect:results}

Mean computation times over the 18 scenarios are given in Table
\ref{tab:results_times}, these show that GAIN3 was the slowest, running in 39 minutes; MALA was next, running in 20 minutes;
followed by GAIN2, in around 4 minutes; and GAIN1, which ran in around 1.5 minutes. These timings are based on the version of the \verb=INLA= package as was available on 8$^{th}$ February 2012, we have since been informed by the authors that recent work on the \verb=laplace= strategy could further reduce computation time by 30--50\%.

\begin{table}[htbp]
    \centering
    \begin{tabular}{ccccc}
          & MALA & GAIN1 & GAIN2 & GAIN3 \\ \hline
        Mean Comp. Time (mins) & 20.236 & 1.443 & 4.292 & 39.147  \\
        Variance & 5.245 & 0.043 & 2.318 & 13.764  \\
    \end{tabular}
    \caption{\label{tab:results_times}Mean and variance of computation time for each algorithm.}
\end{table}

\begin{table}[htbp]
    \centering
    \begin{tabular}{cccccccc}
      Scenario & $\sigma$ & $\phi$ & MALA Last h & \multicolumn{4}{c}{MSE} \\
        &&&&  MALA & GAIN1 & GAIN2 & GAIN3\\ \hline
        1 & 0.5 & 0.02 & 0.099 & 0.211 & 0.214 & 0.213 & 0.212  \\
        3 & 0.5 & 0.04 & 0.104 & 0.175 & 0.185 & 0.185 & 0.188  \\
        5 & 0.5 & 0.06 & 0.103 & 0.179 & 0.175 & 0.175 & 0.173  \\
        7 & 1 & 0.02 & 0.082 & 0.698 & 0.717 & 0.707 & 0.7  \\
        9 & 1 & 0.04 & 0.052 & 0.548 & 0.583 & 0.553 & 0.55  \\
        11 & 1 & 0.06 & 0.049 & 0.443 & 0.481 & 0.442 & 0.435  \\
        13 & 2 & 0.02 & 0.015 & 2.314 & 2.378 & 2.376 & 2.306  \\
        15 & 2 & 0.04 & 0.009 & 1.918 & 2.147 & 1.942 & 1.913  \\
        17 & 2 & 0.06 & 0.004 & 1.737 & 2.371 & 4.382 & 6.792  \\
        &&&&&&&\\
        &&&&&&&\\
   % \begin{tabular}{ccccccc}
   %     Scenario & $\sigma$ & $\phi$ & MALA Last h & MSE MALA & MSE GAIN1 & MSE GAIN2 \\ \hline
        2 & 0.5 & 0.02 & 0.108 & 0.224 & 0.225 & 0.225 & 0.224  \\
        4 & 0.5 & 0.04 & 0.103 & 0.178 & 0.182 & 0.182 & 0.182  \\
        6 & 0.5 & 0.06 & 0.098 & 0.153 & 0.16 & 0.16 & 0.161  \\
        8 & 1 & 0.02 & 0.081 & 0.636 & 0.655 & 0.64 & 0.646  \\
        10 & 1 & 0.04 & 0.066 & 0.56 & 0.598 & 0.563 & 0.559  \\
        12 & 1 & 0.06 & 0.06 & 0.511 & 0.516 & 0.509 & 0.492  \\
        14 & 2 & 0.02 & 0.014 & 2.631 & 2.654 & 2.711 & 2.605  \\
        16 & 2 & 0.04 & 0.013 & 1.954 & 2.078 & 2.053 & 1.937  \\
        18 & 2 & 0.06 & 0.003 & 1.867 & 2.306 & 4.091 & 6.241  \\
    \end{tabular}
    \caption{\label{tab:MSEs}Simulation results: dataset parameters, last value of $h$ from MALA and mean square errors in
predicting the true field. The upper half of the
    table gives results for fixed spatial $\lambda_1(s)$,
    the lower half results for fixed spatial $\lambda_2(s)$.}
\end{table}

Two measures were used to compare the performance of each method: the mean square error in estimating the latent field and a
measure based on estimating probabilities. Details of the parameters for each scenario, the last value of $h$ in the MALA run, and
mean square errors are shown in Table \ref{tab:MSEs}. The mean square error was calculated as,
\begin{equation*}
   \text{MSE} = \frac1{M_{\text{in}}}\sum_s [Y(s) - \hat Y(s)]^2,
\end{equation*}
where $M_{\text{in}}$ is the number of cells inside the observation window and the summation takes place over these cells. The
results show that MALA gave marginally better point-wise prediction of the mean field. With the exception of scenarios 17 and 18, there is a tendency for
INLA with a higher order neighbourhood structure to perform better, though the difference
between the two implementations was only slight. These MSE's are presented to give the reader a sense of the increasing difficulty
of the datasets. They cannot be used for comparing the algorithms because they contain no measure of estimated uncertainty
in the latent field. In order to account for this uncertainty, we now introduce a measure of predictive ability. For each cell $s$ inside the
observation window, both MALA and INLA can produce a set of  estimated quantiles, in this case for probabilities $q_k\in
q=\{0.01,0.05,0.1,0.2,0.3,0.4,0.5,0.6,0.7,0.8,0.9,0.95,0.99\}$, that is a to say a set of thresholds, $c_k(s)$, satisfying,
\begin{equation*}
   \P[Y(s)\leq c_k(s) | X] = q_k.
\end{equation*}
Each algorithm yields a different set of estimated thresholds for each cell, say $c^{(1)}(s)=\{c_k^{(1)}(s)\}$,
$c^{(2)}(s)=\{c_k^{(2)}(s)\}$ and $c^{(3)}(s)=\{c_k^{(3)}(s)\}$ respectively for MALA, GAIN1 and
GAIN2. Now, for $l\in\{1,2,3\}$ we define
\begin{equation*}
   Z_k^{(l)}(s) = \indic[Y(s)\leq c_k^{(l)}(s)],
\end{equation*}
where $\indic$ is the indicator function. Then, $Z_k^{(l)}(s)$ is the indicator function of whether or not the true field in cell $s$
was below the inferred threshold, $c_k^{(l)}(s)$, from method $l$. If $c_k^{(l)}(s)$ is an unbiased estimator of the true
threshold, $c_k^{\text{true}}(s)$, then $\E[Z_k^{(l)}(s)]=q_k$, since,
\begin{equation*}
   q_k = \P[Y(s)\leq c_k^{\text{true}}(s)] = \E\left\{\P[Y(s)\leq c_k^{(l)}(s) | X]\right\} = \E\left\{\E\{\indic[Y(s)\leq
c_k^{(l)}(s)] | X\}\right\}=\E[Z_k^{(l)}(s)].
\end{equation*}
The two measures of predictive ability are bias and mean squared error in estimating the probabilities, $q_k$, over the whole
observation window. The `predictive mean square error' was computed as,
\begin{equation*}
   \text{MSE}_2 = \frac1{13}\sum_{k=1}^{13}\left\{\frac1{M_{\text{in}}}\sum_s [q_k - Z_k^{(l)}(s)]^2\right\}.
\end{equation*}
The results are given in Table \ref{tab:estprobabilities}.

\begin{table}[htbp]
    \centering
    \begin{minipage}{0.5\textwidth}
    \begin{center}
    \begin{tabular}{cccc}
        Scenario &  GAIN1 & GAIN2 & GAIN3\\ \hline
        $1^\star$ & 2.697 & 2.258 & 1.594  \\
        $3^\star$ & 2.801 & 2.748 & 3.553  \\
        5 & 0.645 & 0.646 & 0.53  \\
        $7^\star$ & 4.239 & 4.688 & 0.762  \\
        $9^\star$ & 3.309 & 1.981 & 0.91  \\
        $11^\star$ & 1.156 & 1.045 & 0.614  \\
        $13^\star$ & 26.929 & 82.41 & 0.531  \\
        $15^\star$ & 84.014 & 60.145 & 0.558  \\
        $17^\star$  & 3.907 & 8.796 & 17.63  \\
    \end{tabular}
    \end{center}
    \end{minipage}\begin{minipage}{0.5\textwidth}
    \begin{center}
    \begin{tabular}{cccc}
	Scenario &  GAIN1 & GAIN2 & GAIN3 \\ \hline
        $2^\star$ & 14.192 & 10.666 & 11.747  \\
        $4^\star$ & 6.454 & 6.499 & 7.065  \\
        $6^\star$ & 3.149 & 3.15 & 3.719  \\
        $8^\star$ & 4.077 & 1.49 & 6.719  \\
        $10^\star$ & 2.449 & 1.34 & 0.471  \\
        12 & 0.374 & 0.777 & 0.434  \\
        $14^\star$ & 2.721 & 9.693 & 0.692  \\
        16 & 0.457 & 3.62 & 1.131  \\
        $18^\star$ & 36.912 & 135.375 & 311.099  \\
    \end{tabular}
    \end{center}
    \end{minipage}
    \caption{\label{tab:estprobabilities}Mean square error in estimating probabilities, MSE$_2$, using each of the three INLA
approximations,
    relative to MSE$_2$ for the MALA algorithm. A $^\star$ in the first column indicates the scenarios where MALA outperformed
GAIN1 and GAIN2. The left table are the results for fixed spatial $\lambda_1(s)$ and the right table gives the values for fixed spatial
$\lambda_2(s)$.}
\end{table}

These results show that GAIN1 and GAIN2 outperformed MCMC in only two of the scenarios considered (5, 12 and 16). In the remaining scenarios, the relative mean squared error of GAIN2 varied from 1.045 to 82.41, the median being 3.38. Increasing the neighbourhood size from 1 to 2 did not lead to a median decrease in relative MSE, computed as the median of $\text{MSE}_2(\text{GAIN2})/\text{MSE}_2(\text{GAIN1})$. The median decrease in relative MSE comparing MALA to GAIN3 was 1 (the median of $\text{MSE}_2(\text{GAIN3})/\text{MSE}_2(\text{MALA})$), although GAIN3 performed better in the scenarios with fixed spatial $\lambda_1(s)$ compared with those with $\lambda_2(s)$, with respective values of 0.76 and 3.72.

\begin{figure}[htbp]
   \centering
   \includegraphics[width=0.3\textwidth,height=0.3\textwidth]{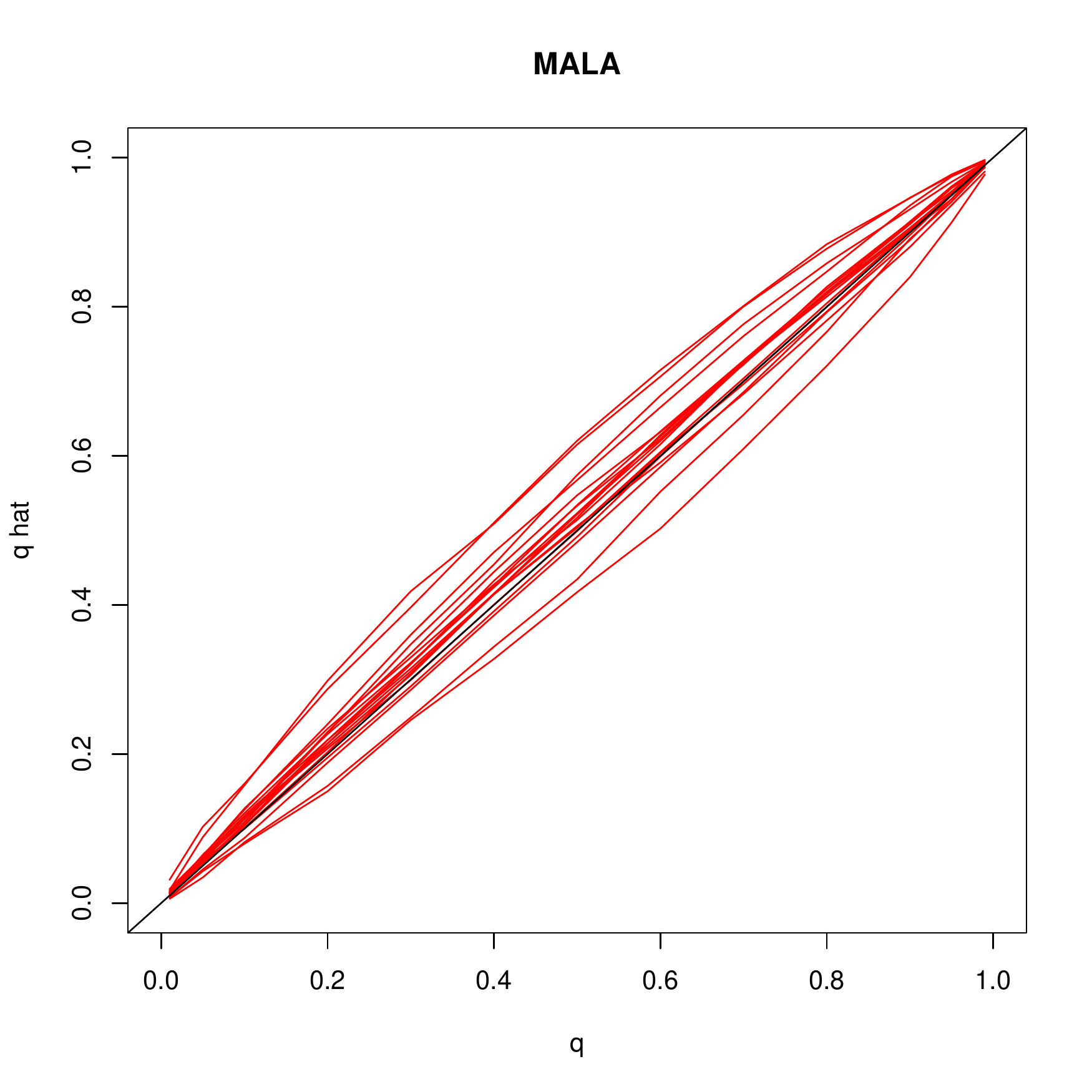}
   \includegraphics[width=0.3\textwidth,height=0.3\textwidth]{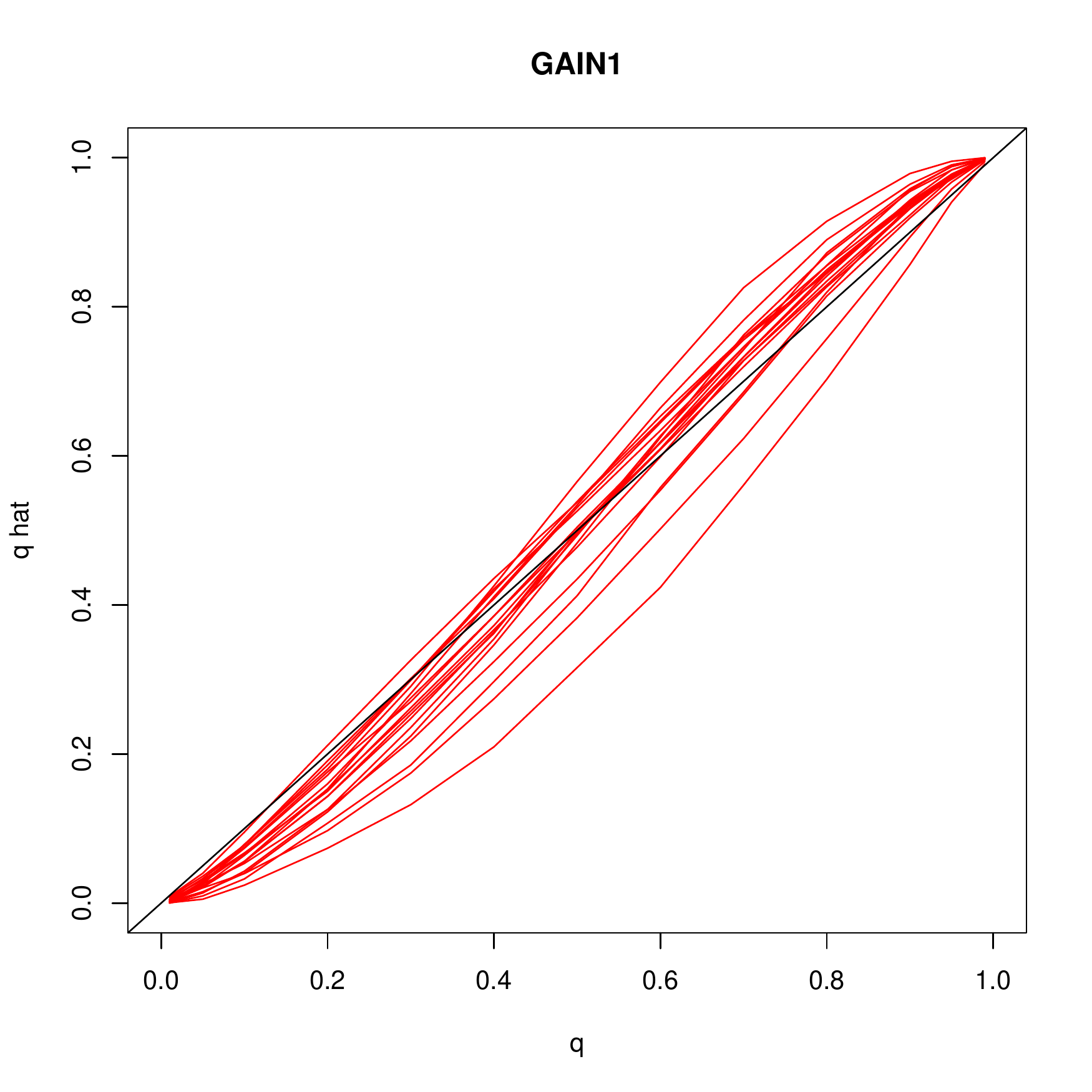}
   
   \includegraphics[width=0.3\textwidth,height=0.3\textwidth]{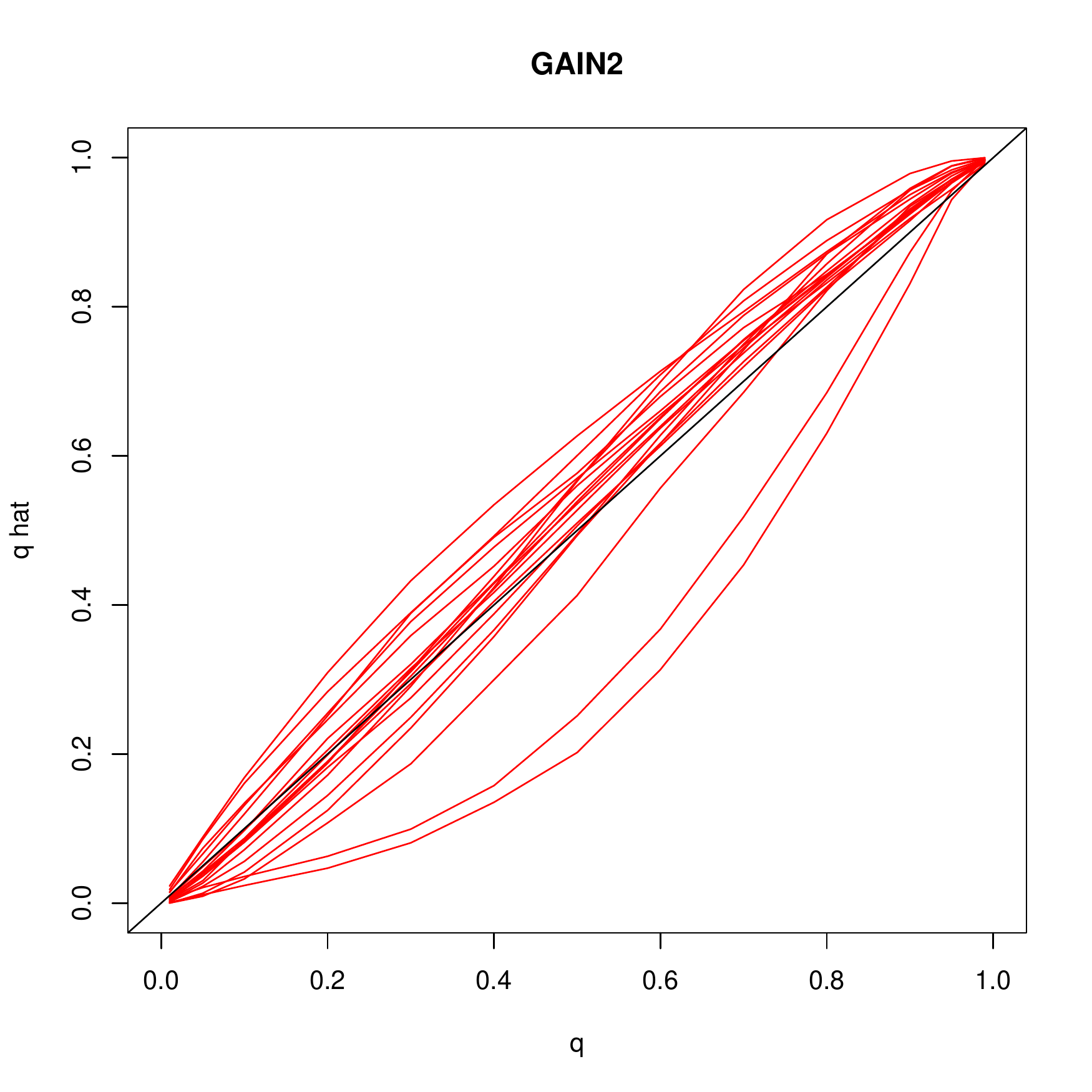}
   \includegraphics[width=0.3\textwidth,height=0.3\textwidth]{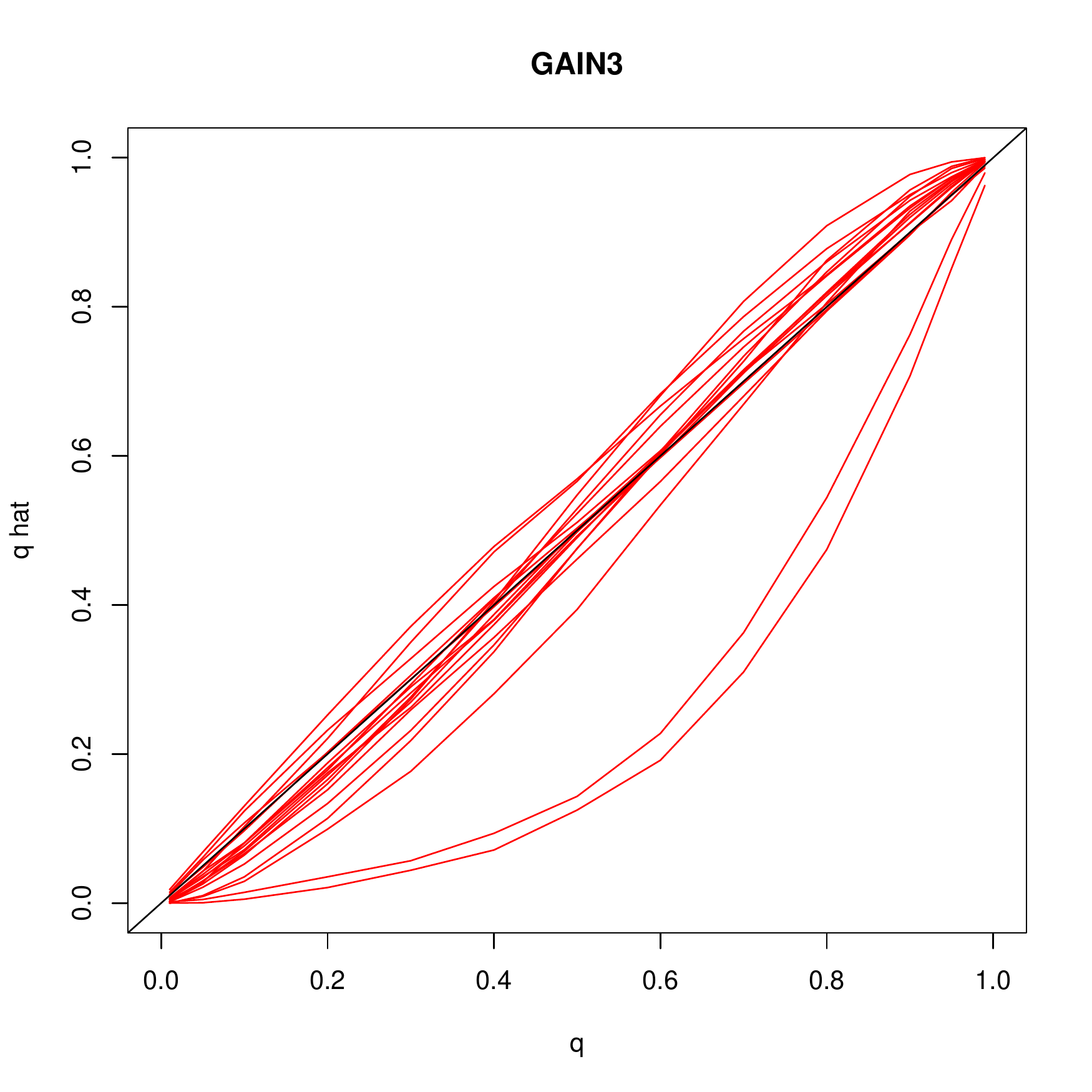}
   \caption{\label{fig:inlavsmalaQt} Illustrating the estimation of quantiles $q_k$ by those inferred by each algorithm $\hat
q_k=\E[Z_k^{(l)}(s)]$ in scenarios 1--18. MALA in the top left plot, GAIN1 in the top right plot and GAIN2 in the bottom left plot and GAIN3 in the bottom right hand plot. For the MALA plot, the estimated quantiles are distributed about the line `$y=x$', whereas for the INLA-based methods, these fall in an `S' shape around that line.}
\end{figure}

The main source of these observed differences in predictive mean squared error is bias in the estimated
probabilities. Figure \ref{fig:inlavsmalaQt} shows plots of estimated versus true probabilities in each of the 18 scenarios for each of the
algorithms. The bias in these plots over the scenarios for both INLA algorithms is apparent in the `S' shape around the line $y=x$, whereas for MCMC, the results are
approximately symmetrically distributed about it. Additionally, these plots show that in two of the scenarios, both GAIN2 and GAIN3 would not give good estimates of probabilities.

% % Bias was computed for each estimated probability as,
% % \begin{equation*}
% %    \text{bias}_k=\frac1{M_{\text{in}}}\sum_s [q_k - Z_k^{(l)}(s)].
% % \end{equation*}
% % Figure \ref{fig:inlavsmalaBiQt} shows density plots of the mean bias (over all probabilities) in each scenario; the means over
% % all scenarios are shown as vertical lines.

\begin{figure}[htbp]
   \centering

   \includegraphics[width=0.3\textwidth,height=0.3\textwidth]{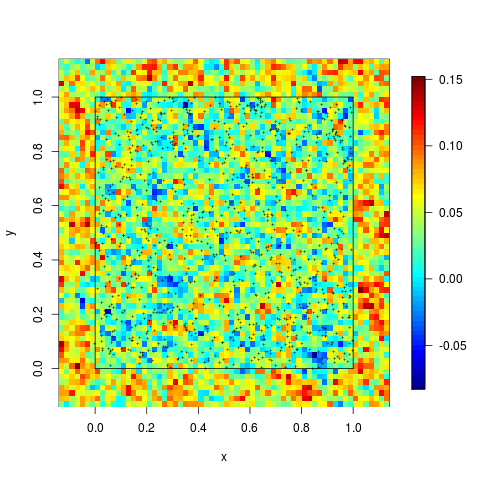}
   \includegraphics[width=0.3\textwidth,height=0.3\textwidth]{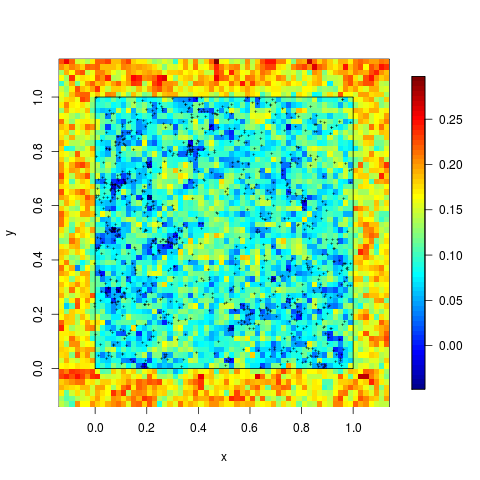}
   \includegraphics[width=0.3\textwidth,height=0.3\textwidth]{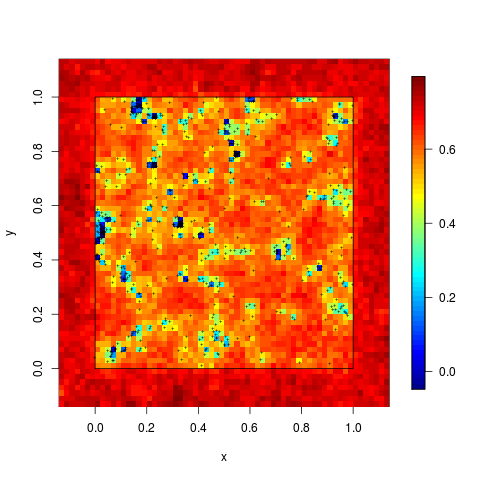}
   \caption{\label{fig:convergence} Image plots of lag-1 autocorrelation of the MCMC chain; the plots correspond to scenarios 4
(left), 10 (middle) and 16 (right). The plots for scenarios in respective blocks 1--6, 7--12 and 13--16 are similar in appearance.
The box in the centre of the plot is the observation window and the data themselves appear as points on each plot.}
\end{figure}

Our convergence diagnostics for MALA consisted of examining the cell-wise lag-1 autocorrelation of the MCMC chain. Example plots are
shown in Figure \ref{fig:convergence}; these were typical of the images produced in the blocks of scenarios 1--6, 7--12 and
13--18. The main feature of these plots is that in every case, the chain is mixing better (i.e.\ has lower lag-1 autocorrelation) in
event-rich areas. Strictly, one should consider the mixing properties across all cells simultaneously, in which case the
conclusion from this diagnostic would be that we would trust the results from scenarios 1--12, but worry about those from 13--18.
However, due to a judicious choice of initial values for the chain (recall that $\Gamma$ was initially set to zero), the effect of
slow mixing in event-poor areas has less effect on the resulting inference than it might with a different set of initial values.
Unconditionally, $\E(Y)=-\sigma^2/2$, so the chain in event-poor areas (sensibly) stays close to this value. In fact, a similar
phenomenon occurs in in INLA: in event-poor areas, the prediction surface tends to $-\sigma^2/2$. Despite the apparently slow
mixing in scenarios 13--18, MALA still produces better predictive inference in each of these cases.

To investigate whether the results for INLA were due to the approximation of the latent field, or whether they were due to error induced in the inference, we conducted a further simulation study. For each of the covariance functions defined by the parameter combinations in scenarios 1--18 above, we constructed a 2-neighbourhood GMRF approximation to the full covariance, and then simulated 18 further scenarios ($1'$--$18'$) based on Gaussian variables simulated from the GMRF. In the following, INLA2 and INLA3 employ the same \verb=INLA= settings as GAIN2 and GAIN3, but in this case there is no GMRF approximation step

\begin{table}[htbp]
    \centering
    \begin{minipage}{0.5\textwidth}
    \begin{center}
    \begin{tabular}{ccccc}
         Scenario & INLA2 & INLA3 \\ \hline
        $1'$ & 0.207 & 0.8  \\
        $3'^\star$ & 8.146 & 0.637  \\
        $5'$ & 0.462 & 1.085  \\
        $7'^\star$ & 6.506 & 0.726  \\
        $9'^\star$ & 1.773 & 1.759  \\
        $11'^\star$ & 14.542 & 0.632  \\
        $13'^\star$ & 167.657 & 3.584  \\
        $15'^\star$ & 2.115 & 2.278  \\
        $17'^\star$ & 2.708 & 0.874  \\
    \end{tabular}
    \end{center}
    \end{minipage}\begin{minipage}{0.5\textwidth}
    \begin{center}
    \begin{tabular}{ccccc}
         Scenario & INLA2 & INLA3 \\ \hline
        $2'^\star$ & 1.922 & 0.913  \\
        $4'^\star$ & 3.381 & 0.933  \\
        $6'$ & 0.344 & 1.141  \\
        $8'^\star$ & 15.142 & 0.622  \\
        $10'^\star$ & 55.562 & 1.743  \\
        $12'$ & 0.032 & 1.365  \\
        $14'^\star$ & 40.266 & 6.144  \\
        $16'^\star$ & 56.126 & 2.5  \\
        $18'$ & 0.545 & 2.586  \\
    \end{tabular}
    \end{center}
    \end{minipage}
    \label{tab:estprobabilitiesGMRF}Mean square error in estimating probabilities, MSE$_2$, using each of the three INLA approximations, relative to MSE$_2$ for the MALA algorithm. A $^\star$ in the first column indicates the scenarios where MALA outperformed INLA2. The left table are the results for fixed spatial $\lambda_1(s)$ and the right table gives the values for fixed spatial $\lambda_2(s)$.
\end{table}

\begin{figure}[htbp]
   \centering
   \includegraphics[width=0.3\textwidth,height=0.3\textwidth]{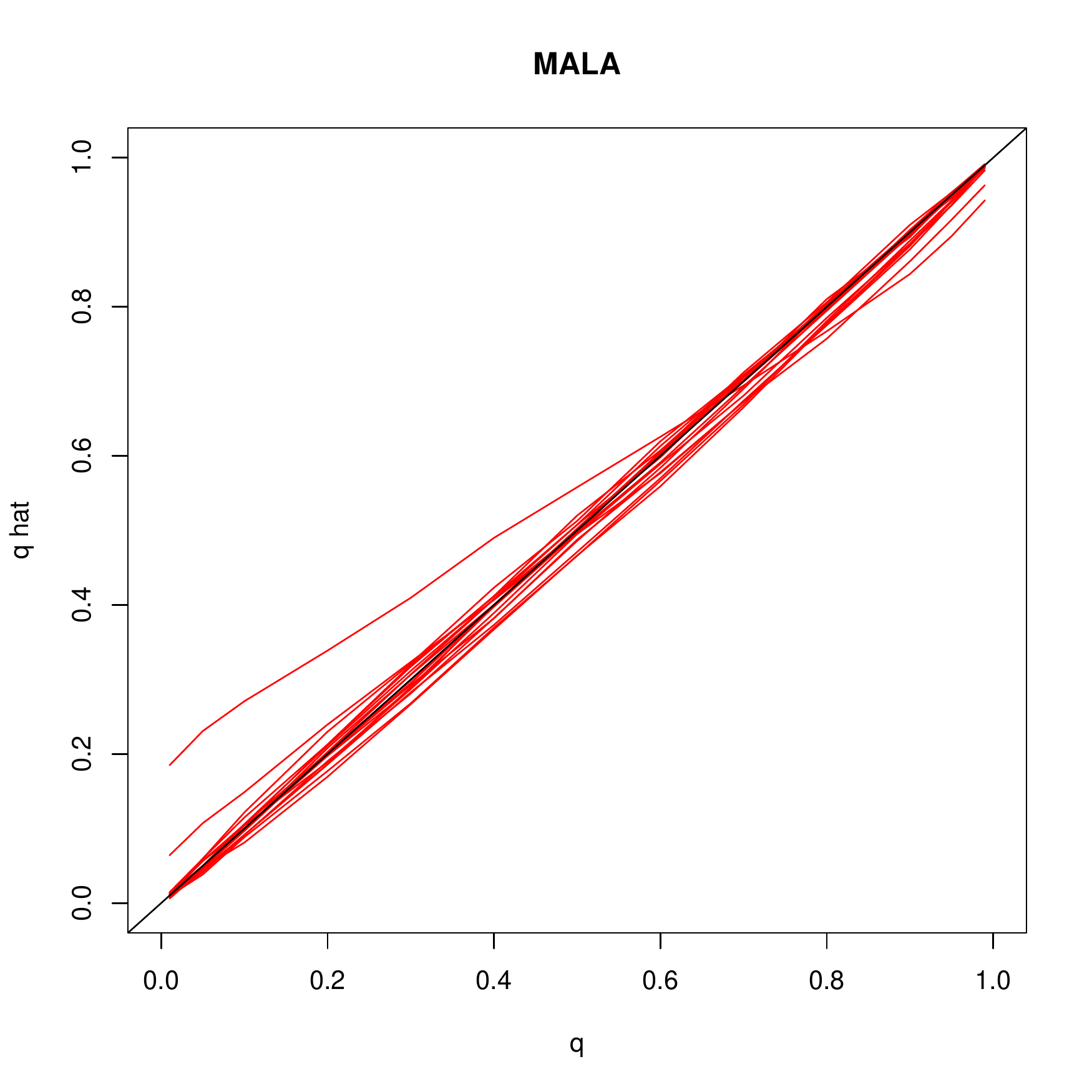}
   \includegraphics[width=0.3\textwidth,height=0.3\textwidth]{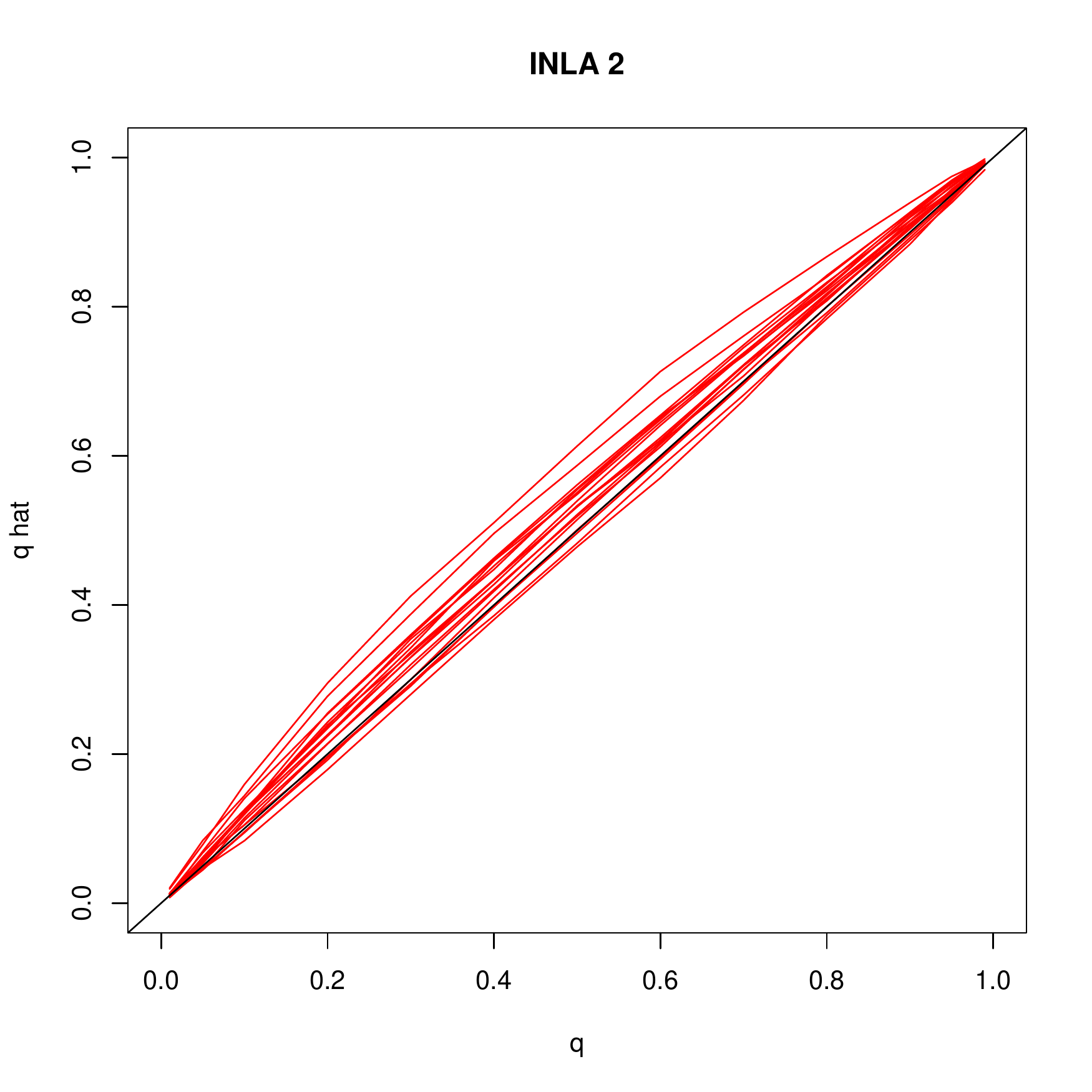}
   \includegraphics[width=0.3\textwidth,height=0.3\textwidth]{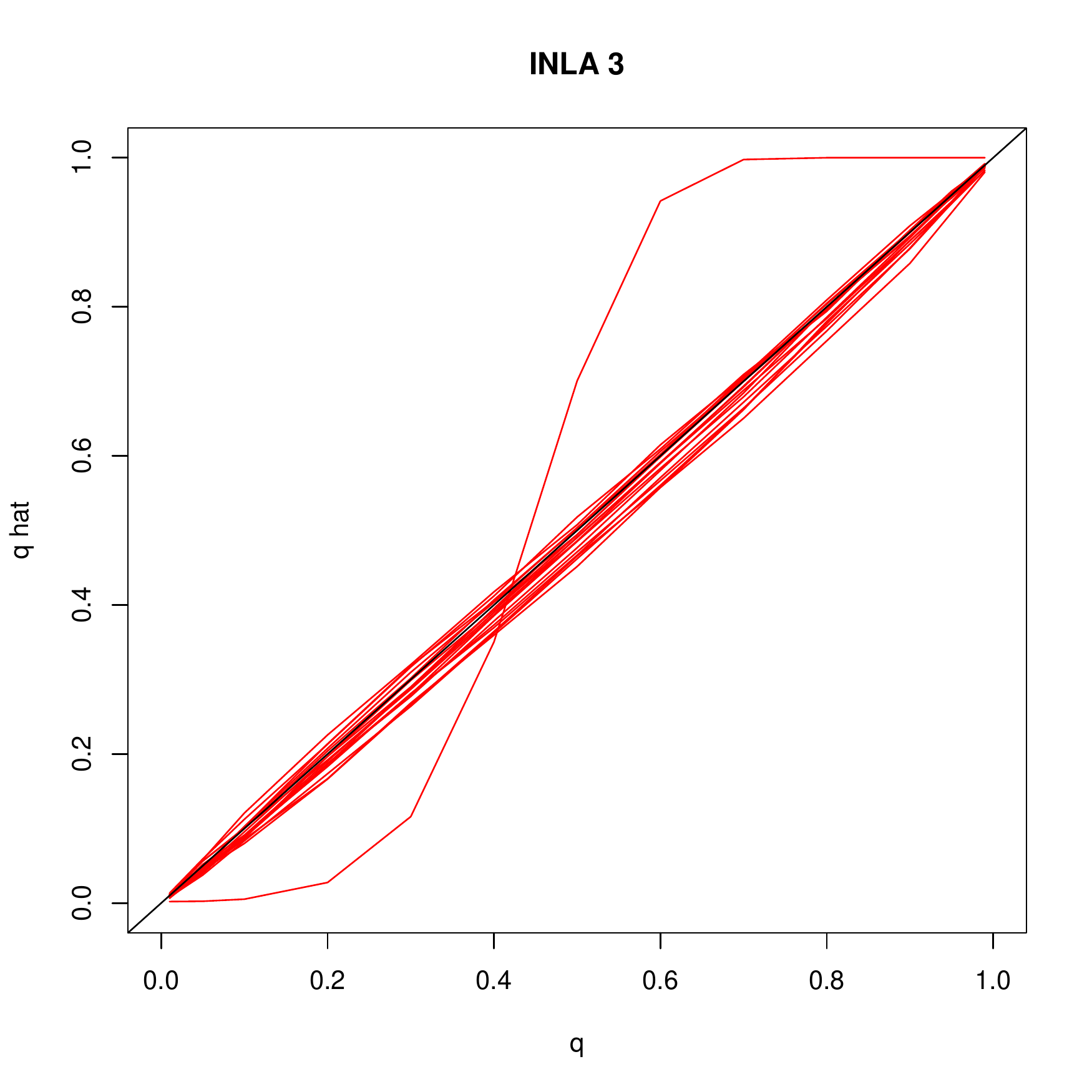}
   \caption{\label{fig:inlavsmalaQtGMRF} Illustrating the estimation of quantiles $q_k$ by those inferred by each algorithm $\hat
q_k=\E[Z_k^{(l)}(s)]$ in scenarios $1'$--$18'$. MALA in left plot, GAIN2 in the middle plot and GAIN3 in the right hand plot.}
\end{figure}

Results from the second simulation study are shown in Table \ref{tab:estprobabilitiesGMRF} and Figure \ref{fig:inlavsmalaQtGMRF}. There are two main conclusions to be drawn from these results. Firstly, in Figure \ref{fig:inlavsmalaQtGMRF}, the `S' shape is no longer apparent for the INLA methods, however in the plot for INLA2, there is a noticeable upward bias in the centre of the plot. Also from the plots, it is clear that MALA did not perform well on two occasions, and INLA on one. The main second conclusion from this simulation study is despite the fact that INLA now shows less bias, MALA nevertheless still outperforms both INLA2 and INLA3. The median relative increase in MSE comparing MALA to INLA2 was 3 over all scenarios and 1.1 comparing MALA to INLA3. As was the case for scenarios 1--18, MALA performs better for fixed spatial $\lambda_2(s)$, with a median increase of 1.37 for INLA3 whereas for $\lambda_1(s)$, INLA3 performed better at a median increase of 0.87.

% > median(res[,3]/res[,1])
% [1] 3.044762
% > median(res[,4]/res[,1])
% [1] 1.113418
% > median(res[(1:18)%%2==1,3]/res[(1:18)%%2==1,1])
% [1] 2.708038
% > median(res[(1:18)%%2==0,3]/res[(1:18)%%2==0,1])
% [1] 3.381486
% > median(res[(1:18)%%2==1,4]/res[(1:18)%%2==1,1])
% [1] 0.874443
% > median(res[(1:18)%%2==0,4]/res[(1:18)%%2==0,1])
% [1] 1.36521

\section{Discussion}
\label{sect:discussion}

In this article, we have provided a tour of the mathematical and statistical techniques behind spatial prediction for log-Gaussian Cox
processes. We have independently evaluated a previously published method for approximating spatially continuous Gaussian processes
with GMRF and conducted a critical comparison of two methodologies for predictive Bayesian inference in this class of models.

A suite of functions (as well as wrapper functions for the approximate Bayesian predictive inference for INLA) have
been made freely available in the \verb=lgcp= R package \citep{taylor2011a}. Our restriction in this paper to spatial, rather than spatio-temporal prediction is primarily ease of exposition. However, and unsurprisingly, the spatio-temporal concept is computationally more demanding. In pilot runs, even INLA was found to be quite slow for spatio-temporal prediction on a regular grid, due to the hugely increased dimensionality of the problem and a corresponding increase in the complexity of dependence patterns in the precision
matrix (see Section \ref{sect:gmrfapprox}). Furthermore, we have restricted our choice of MCMC methods to the Metropolis
adjusted Langevin algorithm (MALA) rather than investigating more sophisticated sampling techniques such as Riemann manifold
Langevin or Hamiltonian Monte Carlo \citep{girolami2011}. Our reasons for this choice are ease of implementation, stability and
the fact that MALA has been well studied in the literature, with various theoretical results available concerning practical implementation. The authors are aware that the methods of \cite{girolami2011} have better theoretical mixing properties.

We have demonstrated that MCMC can yield more accurate estimates of predictive probabilities compared with INLA-based methods for this class of models, depending on the chosen settings. Furthermore the predictive probabilities from MCMC can be obtained comparatively quickly, and show less bias compared with those from INLA. The inferential technique of producing a GMRF approximation to a Gaussian field and then performing inference via the integrated nested Laplace approximation should therefore be regarded with appropriate caution. This article also opens up the question of the utility of gradient-based MCMC methods for inference in log-Gaussian Cox process models assuming a latent Gaussian Markov random field; the default method of inference for these models would appear to be a blocked Gibbs sampling strategy eg.\ \cite{rue2009}.

We have not addressed the full capabilities of both software implementations: \verb=lgcp= and \verb=INLA=. In particular,
\verb=INLA= provides access to inference for a wide class of latent Gaussian models, whilst \verb=lgcp= is restricted to spatial
and spatio-temporal log-Gaussian Cox processes. Furthermore, the \verb=INLA= package also provides a framework for the estimation of hyperparameters, which \verb=lgcp= does not; this is known to be a very challenging sampling task for MCMC.

\end{document}